\documentclass{amsart}
\usepackage{amsfonts,amssymb}
\usepackage{amsmath}
 \makeatletter\@addtoreset{equation}{section}\makeatother

\newtheorem{theorem}{Theorem}[section]
\newtheorem{corollary}[theorem]{Corollary}
\newtheorem{lemma}[theorem]{Lemma}
\newtheorem{proposition}[theorem]{Proposition}
\theoremstyle{definition}

\theoremstyle{remark}

\numberwithin{equation}{section}

\begin{document}

\title[Hierarchical Model of Quantum Anharmonic Oscillators ]
{A Hierarchical Model of Quantum Anharmonic Oscillators: Critical Point Convergence}%

\author{S. Albeverio}
\address{Abteilung f\"ur Stochastik, Universit\"at Bonn,
        D 53115 Bonn,
        Germany\\ BiBoS Research Centre, Bielefeld, Germany \\ CERFIM, Locarno and Academia di Architettura,\\
 Universit\`a della Svizzera Italiana, Mendrisio,  Switzerland\\ Dipartimento di Matematica,
 Universit\`a di Trento, Trento, Italy}
\email{albeverio@uni-bonn.de}
\author{Yu. Kondratiev}
\address{Fakult\"at f\"ur Mathematik, Universit\"at Bielefeld, D 33615 Bielefeld, Germany\\
BiBoS Research Centre, Bielefeld, Germany\\ Institute of
Mathematics, Kiev, Ukraine  }
\email{kondrat@mathematik.uni-bielefeld.de}
\author{A. Kozak}
\address{Instytut  Matematyki, Uniwersytet Marii Curie-Sk{\l}odowskiej \\
        PL 20-031 Lublin, Poland}
  \email{akozak@golem.umcs.lublin.pl}
  \author{Yu. Kozitsky}
  \address{Instytut  Matematyki, Uniwersytet Marii Curie-Sk{\l}odowskiej \\
        PL 20-031 Lublin, Poland}
  \email{jkozi@golem.umcs.lublin.pl}

\thanks{Supported by Deutsche Forschungsgemeinschaft through the German-Polish project
436 POL 113/98/0-1 ``Probability measures"; Agnieszka Kozak and Yuri Kozitsky were also supported by Komitet Bada{\'n}
Naukowych through the Grant 2P03A 02025. }%
\subjclass{82B27; 82B10}%
\keywords{Euclidean Gibbs states; abnormal fluctuations; criticality; quantum crystal}%
\dedicatory{to appear in Communications in Mathematical Physics}
\begin{abstract}
A hierarchical model of interacting quantum particles performing
 anharmonic oscillations
is studied in the Euclidean approach, in which the local Gibbs states are constructed as measures on infinite
dimensional spaces. The local states restricted to the subalgebra generated by fluctuations of displacements of
particles are in the center of the study. They are described by means of the corresponding temperature Green
(Matsubara) functions.  The result of the paper is a theorem, which describes the critical point convergence of such
Matsubara functions in the thermodynamic limit.
\end{abstract}
\maketitle

\tableofcontents

\section{Introduction}

Let $\mathbb{L}$ be a countable set (lattice). With each $l
\in\mathbb{L}$ we associate a quantum mechanical particle with one
degree of freedom described by the momentum $\mathfrak{p}_l$ and
 displacement $\mathfrak{q}_l$ operators. The system of
such particles which we consider in this article is described by the heuristic Hamiltonian
\begin{equation} \label{au1}
H = -\frac{1}{2}\sum_{l, l'} J_{ll'} \mathfrak{q}_l
\mathfrak{q}_{l'} + \sum_{l}
\left[\frac{1}{2\mathfrak{m}}\mathfrak{p}_l^2 + \frac{a}{2}
\mathfrak{q}_l^2 + {b} \mathfrak{q}_l^4  \right].
\end{equation}
Here $b>0$, $a\in \mathbb{R}$ and the sums run through the lattice $\mathbb{L}$. The operators $\mathfrak{p}_l$ and
$\mathfrak{q}_l$ satisfy the relation
\begin{equation} \label{au2}
[\mathfrak{p}_l, \mathfrak{q}_l ] =  \mathfrak{p}_l \mathfrak{q}_l - \mathfrak{q}_l \mathfrak{p}_l = 1/i,
\end{equation}
and $\mathfrak{m} =\mathfrak{m}^{\rm phys}/ \hbar^2$ is the reduced mass of the particle. Models like (\ref{au1}) have
been studied for many years, see e.g., \cite{pk,stamen}. They (and their simplified versions) are used as a base of
models describing strong electron-electron correlations caused by the interaction of electrons with vibrating ions
\cite{freer,stas1}.

Let $\mathcal{L} =\{\Lambda_n\}_{n \in \mathbb{N}_0}$, $\mathbb{N}_0 = \mathbb{N} \cup \{0\}$ be a sequence of finite
subsets of $\mathbb{L}$, which is ordered by inclusion and exhausts $\mathbb{L}$. For every $\Lambda_n$, let
$H_{\Lambda_n}$ be a local Hamiltonian, corresponding to (\ref{au1}). In a standard way the Hamiltonians
$H_{\Lambda_n}$ determine local Gibbs states $\varrho_{\beta, \Lambda_n}$. A phase transition in the model (\ref{au1})
is connected with macroscopic displacements of particles from their equilibrium positions $\mathfrak{q}_l = 0$, $l \in
\mathbb{L}$. To describe this phenomenon, one considers fluctuation operators
\begin{equation} \label{au3}
Q_{\Lambda_n}^{(\alpha)} \ \stackrel{\rm def}{=} \
\frac{1}{|\Lambda_n|^{(1+\alpha)/2}} \sum_{l\in \Lambda_n }
\mathfrak{q}_l, \quad \alpha \geq 0,
\end{equation}
and Matsubara functions
\begin{eqnarray} \label{au4}
& & \Gamma_{2k}^{\alpha, \beta , \Lambda_n}(\tau_1 , \dots,
\tau_{2k}) \ \stackrel{\rm def}{=} \ \varrho_{\beta , \Lambda_n}
\left\{Q_{\Lambda_n}^{(\alpha)} \exp\left(-(\tau_2 -
\tau_1)H_{\Lambda_n}\right) \cdots \right. \\ & &\quad  \times
\left. \exp\left(-(\tau_{2k} -
\tau_{2k-1})H_{\Lambda_n}\right)Q_{\Lambda_n}^{(\alpha)}
\exp\left((\tau_{2k} - \tau_{1})H_{\Lambda_n}\right)\right\},
\quad k \in \mathbb{N}, \nonumber
\end{eqnarray}
with the arguments satisfying the condition $0 \leq \tau_1 \leq
\dots \leq \tau_{2k} \leq \beta$. In our model the interaction
potential is taken to be
\begin{equation} \label{ip}
J_{ll'} = J[d(l,l') +1]^{-1 - \delta}, \quad J , \delta >0,
\end{equation}
where $d(l,l')$ is a metric on $\mathbb{L}$, determined by means of a hierarchical structure. The latter is a family of
finite subsets of $\mathbb{L}$, each of which belongs to a certain hierarchy level $n \in \mathbb{N}_0$. This fact
predetermines also our choice of the sequence $\mathcal{L}$ -- the subsets $\Lambda_n$ are to be the elements of the
hierarchical structure, that is typical for proving scaling limits in hierarchical models (see e.g., \cite{BM}). We
prove (Theorem \ref{h1tm}) that, for any $\delta \in (0, 1/2)$, the parameters $a\in \mathbb{R}$, ${b}
> 0$ and $\mathfrak{m}>0$ can be chosen in such a way that there
will exist $\beta_* >0$ with the following properties: \vskip.1cm
\begin{tabular}{ll}
{\rm (a)} &if $\beta = \beta_*$, for all $k \in \mathbb{N}$,  the functions (\ref{au4}) converge
\end{tabular} \vskip.1cm
\begin{equation} \label{au5}
\Gamma_{2k}^{\delta, \beta_* , \Lambda_n}(\tau_1 , \dots,
\tau_{2k}) \longrightarrow \frac{(2k)!}{k! 2^k
\beta_*^k}\left(\frac{J_*}{J} \right)^k, \quad n \rightarrow
+\infty
\end{equation}
\vskip.1cm
\begin{tabular}{ll}
&uniformly with respect to their arguments; here $J_*>0 $ is a costant\\ &determined by the hierarchical structure;
\\[.2cm] {\rm (b)} &if $\beta < \beta_*$, for all $\alpha >0$ and $k \in
\mathbb{N}$, the functions $\Gamma_{2k}^{\alpha, \beta ,
\Lambda_n}$ converge\\ &to zero in the same sense.
\end{tabular}
\vskip.1cm \noindent The convergence of the functions (\ref{au4})
like in (\ref{au5}) but with $\alpha = 1$ would correspond to the appearance of a long-range order, which destroys the
$Z_2$-symmetry. Thus, claim (a) describes a critical point where the
 fluctuations are abnormal (since $\alpha = \delta
>0$) but not strong enough to destroy the mentioned symmetry.
Such fluctuations are classical (non-quantum), which follows from the fact that the limits (\ref{au5}) are independent
of $\tau$.

Due to the hierarchical structure the model (\ref{au1}) is self-similar. In translation invariant lattice models
self-similarity appears at their critical points
\cite{Sinai1,Sinai}. This, among others, is the reason why the
critical point properties of hierarchical models of classical statistical mechanics attract attention during the last
three decades. An expository review of the results in this domain is given in \cite{BM}.

In the model (\ref{au1}) the oscillations are described by unbounded operators\footnote{Certain aspects of critical
point behaviour of quantum hierarchical models with bounded (spin) operators were studied in \cite{MoSch}.}. The same
model was studied in our previous works \cite{AKK1,AKK2,AKK3}. In
\cite{AKK1} a preliminary study of the model was performed. A
theorem describing the critical point convergence was announced in
\cite{AKK2}. In \cite{AKK3} we have shown that the critical point
of the model (\ref{au1}) can be suppressed by strong quantum effects, which take place, in particular, when the mass
$\mathfrak{m}$ is less than a certain bound $\mathfrak{m}_*$\footnote{Physical aspects of such quantum effects were
analyzed in \cite{prl}.}. In the present paper we give a complete proof of the critical point convergence, which
appears for sufficiently large values of the mass (see the discussion at the very end of this introduction). It should
be pointed out that, to the best of our knowledge, our result is the first example of a theorem, which describes the
convergence at the critical point of a nontrivial quantum model, published by this time.

Let us outline the main aspects of the proof. By symmetry, the functions (\ref{au4}) are extended to
$\mathcal{I}_\beta^{2k}$, where $\mathcal{I}_\beta \ \stackrel{\rm def}{=} \ [0, \beta]$. Then for $x \in L^2
(\mathcal{I}_\beta)$, one sets
\begin{equation} \label{au6}
\varphi_n^{(\alpha)} (x) = 1 + \sum_{k=1}^{\infty}\frac{1}{(2k)!}
\int_{\mathcal{I}_\beta^{2k}} \Gamma_{2k}^{\alpha, \beta ,
\Lambda_n}(\tau_1 , \dots, \tau_{2k}) x(\tau_1) \cdots
x(\tau_{2k}) {\rm d}\tau_1 \cdots {\rm d}\tau_{2k},
\end{equation}
and
\begin{equation} \label{au7}
\log \varphi_n^{(\alpha)} (x) = \sum_{k=1}^{\infty}\frac{1}{(2k)!}
\int_{\mathcal{I}_\beta^{2k}} U_{2k}^{\alpha, \beta ,
\Lambda_n}(\tau_1 , \dots, \tau_{2k}) x(\tau_1) \cdots
x(\tau_{2k}) {\rm d}\tau_1 \cdots {\rm d}\tau_{2k},
\end{equation}
which uniquely determines the Ursell functions $ U_{2k}^{\alpha,
\beta , \Lambda_n}$. In terms of these functions our result may be
formulated as follows:
\begin{eqnarray} \label{au8}
& & \qquad \qquad U_{2}^{\delta, \beta_* , \Lambda_n}(\tau_1 ,
\tau_{2})
 \longrightarrow \beta_* ^{-1}, \\
& & \quad \quad \forall k >1: \  U_{2k}^{\delta, \beta_* ,
\Lambda_n}(\tau_1 , \dots, \tau_{2k}) \longrightarrow 0, \nonumber
\\ & &
 \forall k \in \mathbb{N}, \ \beta < \beta_*, \  \alpha > 0: \
 U_{2k}^{\alpha, \beta , \Lambda_n}(\tau_1 , \dots, \tau_{2k})
\longrightarrow 0, \nonumber
\end{eqnarray}
which holds uniformly with respect to the arguments $\tau_j\in
\mathcal{I}_\beta$, $j = 1, \dots, 2k$ as $n \rightarrow +\infty$.
Here we have set $J = J_*$, that can always be done by choosing an appropriate scale of $\beta$. We prove (\ref{au8})
in the framework of the Euclidean approach in quantum statistical mechanics based on the representation of the
functions (\ref{au4}) in the form of functional integrals. This approach was initiated in \cite{AHK,HK}, its detailed
description and an extended related bibliography may be found in \cite{AKKR}. In separate publications we are going to
exploit our result, in particular, to construct self-similar Gibbs states (in the spirit of \cite{Bleher,BM} where it
 was done for classical
hierarchical models).

The functions $\Gamma_{2k}^{\alpha, \beta , \Lambda_n}$, $U_{2k}^{\alpha, \beta , \Lambda_n}$, $k\in \mathbb{N}$ are
continuous on $\mathcal{I}_\beta^{2k}$, see \cite{AKKR}. In view of our choice of the potential energy in (\ref{au1}),
the Ursell functions satisfy the sign rule
\begin{equation} \label{au9}
(-1)^{k-1}U_{2k}^{\alpha, \beta , \Lambda_n}(\tau_1 , \dots, \tau_{2k}) \geq 0,
\end{equation}
for all $k\in \mathbb{N}$ and $(\tau_1 , \dots , \tau_{2k}) \in
\mathcal{I}_\beta^{2k}$. We prove that the families
$\{\Gamma_{2k}^{\alpha, \beta , \Lambda_n}\}_{n \in
\mathbb{N}_0}$, $\{U_{2k}^{\alpha, \beta , \Lambda_n}\}_{n \in
\mathbb{N}_0}$, $k \in \mathbb{N}$ are equicontinuous; hence, the
convergence (\ref{au8}) can be proven by showing the convergence of $U_{2}^{\delta, \beta_* , \Lambda_n}$,
$U_{2}^{\alpha, \beta ,
\Lambda_n}$, as in (\ref{au8}), and
\begin{equation} \label{au10}
\mathcal{U}_{2k}^{\alpha, \beta , \Lambda_n} \ \stackrel{\rm
def}{=} \ \int_{\mathcal{I}^{2k}_\beta}U_{2k}^{\alpha, \beta ,
\Lambda_n}(\tau_1 , \dots, \tau_{2k}){\rm d}\tau_1 \cdots {\rm
d}\tau_{2k}\longrightarrow 0,
\end{equation}
which has to hold for all $k \geq 2, \ \beta \leq \beta_*$, and for $\alpha = \delta$ if $\beta = \beta_*$, and $\alpha
>0$ if $\beta <\beta_*$. Another fact which we employ here is also a consequence of the choice of the potential energy
in (\ref{au1}). By a version of the Lee-Yang theorem, the function $f_n$ of a single complex variable defined in the
vicinity of $z = 0$ by the series
\begin{equation} \label{au11}
\log f_n (z) = \sum_{k=1}^{\infty} \frac{1}{(2k)!}
\mathcal{U}_{2k}^{\alpha, \beta , \Lambda_n} z^{2k} ,
\end{equation}
can be extended to an even entire function of order less than two possessing imaginary zeros only. This implies
\begin{equation}
\label{au12}
 f_n (z) = \prod_{j =1}^{\infty} ( 1 + c_j^{(n)} z^2), \quad c_1^{(n)}
  \geq c_2^{(n)} \geq \cdots >0, \quad \sum_{j=1}^\infty c_j^{(n)} <\infty,
\end{equation}
yielding for the numbers (\ref{au10}) the following representation
\begin{equation} \label{au13}
\mathcal{U}_{2k}^{\alpha, \beta , \Lambda_n} = 2 (2k-1)!
(-1)^{k-1} \sum_{j=1}^\infty \left[c_j^{(n)}  \right]^{k}, \quad k
\in \mathbb{N},
\end{equation}
by which,
\begin{eqnarray} \label{au14}
|\mathcal{U}_{2k}^{\alpha, \beta , \Lambda_n}|& \leq & 2 (2k-1)!
\left[c_1^{(n)}  \right]^{k-2}\sum_{j=1}^\infty \left[c_j^{(n)}
\right]^{2}, \quad k \geq 2,\\ |\mathcal{U}_{2k}^{\alpha, \beta ,
\Lambda_n}|& \leq & (2k-1)! \left[c_1^{(n)}
\right]^{k-1}\mathcal{U}_{2}^{\alpha, \beta , \Lambda_n} \quad k
\in \mathbb{N}, \nonumber
\end{eqnarray}
and hence
\begin{eqnarray}
\label{au14z} |\mathcal{U}_{2k}^{\alpha, \beta , \Lambda_n}|
  & \leq & (2k-1)! (2^{1-k}/3) \left[ \mathcal{U}_{2}^{\alpha,
\beta , \Lambda_n}\right]^{k-2} |\mathcal{U}_{4}^{\alpha, \beta ,
\Lambda_n}|, \quad k\geq 2 ,\\
|\mathcal{U}_{2k}^{\alpha, \beta , \Lambda_n}| & \leq & (2k-1)! 2^{1-k} \left[ \mathcal{U}_{2}^{\alpha, \beta ,
\Lambda_n}\right]^{k},\quad k \in \mathbb{N}. \nonumber
\end{eqnarray}
Then the convergence (\ref{au8}) follows from the corresponding convergence of ${U}_{2}^{\alpha, \beta , \Lambda_n}$
and from the fact
\begin{equation} \label{au15}
\mathcal{U}_{4}^{\delta, \beta_* , \Lambda_n} \longrightarrow 0.
\end{equation}
 The above arguments allow us to prove the convergence of
an infinite number of sequences of functions by controlling just two sequences of numbers -- $\{\hat{u}_n \}_{n \in
\mathbb{N}_0}$ and $\{\mathcal{U}^{\delta, \beta_* , \Lambda_n}_{4}\}_{n \in
\mathbb{N}_0}$, where $\hat{u}_n  =
\beta^{-1}\mathcal{U}_{2}^{\delta, \beta , \Lambda_n}$. The sign
rule (\ref{au9}) and the representation (\ref{au13}) are proven in Lemmas \ref{5.7lm} and \ref{4lm} below by means of
the lattice approximation technique \cite{AKKR}. Here the functions $\Gamma^{\alpha, \beta, \Lambda_n}_{2k}$, $k\in
\mathbb{N}$, are obtained as limits of moments of Gibbs measures of classical ferromagnetic $\phi^4$-models. This
allows us to employ the corresponding properties of the $\phi^4$-models proven in
\cite{Schl} (the sign rule), \cite{AKK1} (a correlation
inequality) and \cite{LS} (the Lee-Yang theorem). Then to controlling the sequences $\{\hat{u}_n \}_{n \in
\mathbb{N}_0}$ and $\{\mathcal{U}^{\delta, \beta_* , \Lambda_n}_{4}\}_{n \in
\mathbb{N}_0}$ we apply a version of the inductive method
developed in \cite{Koz,KW}. The central role here is played by Lemma \ref{5.5lm}. It establishes the existence of
$\beta_*>0$ such that, for $\beta = \beta_*$ (respectively, for $\beta <
\beta_*$), the sequence $\{\hat{u}_n \}_{n \in \mathbb{N}_0}$
converges to one (respectively, to zero as $|\Lambda_n|^{-\delta}$ ). The sequence $\{\mathcal{U}^{\delta, \beta ,
\Lambda_n}_{4}\}_{n \in \mathbb{N}_0}$ converges to zero in both
cases. The latter fact is proven by constructing a converging to zero sequence of positive numbers $\{X_n\}_{n \in
\mathbb{N}_0}$, such that $\beta^{-2}|\mathcal{U}^{\delta, \beta , \Lambda_n}_{4}| \leq X_n$ for all $n \in
\mathbb{N}_0$ and $\beta \leq \beta_*$. The proof of Lemma \ref{5.5lm} is based on recurrent estimates (Lemma
\ref{6.2lm}) yielding upper and lower bounds for $\hat{u}_{n} $
and $X_n$ in terms of certain functions of $\hat{u}_{n-1}$ and $X_{n-1}$. The analysis of these estimates shows that
the simultaneous convergence $\hat{u}_n  \rightarrow 1$ and $X_n
\rightarrow 0$ can be guaranteed if these sequences are confined
to the intervals $\hat{u}_n  \in (1, \bar{v})$ and $X_n \in (0 ,
\bar{w})$, where the parameters $\bar{v}>1$ and $\bar{w}>0$ depend
on $\delta$ and on the details of the hierarchical structure only and can be computed explicitly. Lemma \ref{6.2lm} is
proven by comparing solutions of certain differential equations, similarly as in
 \cite{Koz,KW}. Lemma
\ref{6.3lm} establishes the existence of $\beta_n^{\pm} >0$,
$\beta_n^{-} < \beta_n^{+}$ if $\beta_{n-1}^{\pm}$
 do exist. These numbers are defined as follows:
 $\hat{u}_n  = \bar{v}$ for $\beta = \beta_n^+$, and
$\hat{u}_n  < \bar{v}$ for $\beta < \beta_n^+$;  $\hat{u}_n = 1$ for $\beta = \beta_n^-$, and $\hat{u}_n < 1$ for
$\beta <
\beta_n^-$. The proof of Lemma \ref{6.3lm} is carried out by means
of the estimates obtained in Lemma \ref{6.2lm}. In Lemma
\ref{osz1lm} we prove that the parameters
$\mathfrak{m}$, $a$ and $b$ can be chosen in such a way that $\beta_0^{\pm}$ do exist. In Lemma \ref{Newlm} we prove
the existence of $\beta_*$, such that $\forall n \in \mathbb{N}_0: \
\hat{u}_n \in (1, \bar{v})$ for $\beta = \beta_*$, and $\hat{u}_n
\rightarrow 0$ as $|\Lambda_n|^{-\delta}$ for $\beta < \beta_*$.
The proof is based on the estimates obtained in Lemma \ref{6.2lm}. In Lemma \ref{ormlm} we prove that all $\hat{u}_n$,
$n\in
\mathbb{N}_0$ are continuous functions of $\beta$ and  describe
certain useful properties of the Ursell functions $U^{\alpha_* ,
\beta , \Lambda_n}_2(\tau, \tau')$, $n \in \mathbb{N}_0$, implying
e.g., the mentioned equicontinuity.

The proof of Lemma \ref{osz1lm} is based on the estimates of $\hat{u}_0 $ and $X_0$ obtained in Lemma \ref{alm}. In
particular, we prove that
\[
\frac{\mathfrak{m} \gamma^2}{36} \left[1 - \exp\left(-
\frac{3\beta}{\mathfrak{m}\gamma}\right) \right]\leq \hat{u}_0
\leq \frac{\beta \gamma}{8}\left[1 + \sqrt{1+ \frac{16}{\beta
\gamma}} \ \right],
\]
where $a<0$, $\gamma = |a|/b$. Then for $\mathfrak{m} \gamma^2
> 36 \bar{v}$, one gets $\hat{u}_0 > \bar{v}$ for sufficiently
large $\beta$. On the other hand, $\hat{u}_0 \rightarrow 0$ as $\beta \rightarrow 0$. Since $\hat{u}_0$ depends on
$\beta$ continuously, this yields the existence of $\beta_0^{\pm}$. Furthermore, for fixed $\gamma$ and $\beta$, we
show that $X_0
\leq b C$ with a certain fixed $C>0$. This was used to provide
$X_0 < \bar{w}$, and hence $X_n < \bar{w}$, $n\in \mathbb{N}$, for sufficiently small $b>0$. Another upper bound of
$\hat{u}_0$ was obtained in \cite{AKK3}.
 It
is well-known that the one particle Hamiltonian which stands in the square brackets in (\ref{au1}) has a pure point
non-degenerate spectrum. Let $E_n$, $n\in \mathbb{N}_0$ be its eigenvalues and $\Delta = \min_{n\in \mathbb{N}}(E_n -
E_{n-1})$. In \cite{AKK3} we proved that if $\mathfrak{m}
\Delta^2 > 1$, then $\hat{u}_0 < 1$ and hence $\hat{u}_n
\rightarrow 0$ for all $\beta$. In what follows, the critical
point of the model exists if $a<0$ and the parameters $\mathfrak{m} (|a|/b)^2$, $1/b$ are big enough; such a point does
not exist if `the quantum rigidity' $\mathfrak{m}\Delta^2$ (see
\cite{prl}) is greater than 1. By Lemma 1.1 of \cite{AKK3},
$\mathfrak{m}\Delta^2 \sim \mathfrak{m}^{-1/3} C $, $C>0$ as $\mathfrak{m} \rightarrow 0$, which means that small
values of the mass prevent the system from criticality.

\section{Setup and the Theorem}

Like in \cite{AKK2,AKK3} we consider the hierarchical model defined on $\mathbb{L} = \mathbb{N}_0 $. Given $\varkappa
\in
\mathbb{N}\setminus \{1\}$, we set
\begin{equation} \label{h1}
\Lambda_{n, s} = \{l\in \mathbb{N}_0 \ | \ \varkappa^n s \leq l \leq \varkappa^n
(s+1) -1 \}, \quad s , n \in \mathbb{N}_0.
\end{equation}
Then for $n \in \mathbb{N}$, one has
\begin{equation} \label{h2}
\Lambda_{n, s} = \bigcup_{l \in \Lambda_{k, s}}\Lambda_{n - k, l},
\quad k = 1, 2, \dots , n .
\end{equation}
The collection of families $\{ \Lambda_{n, s}\}_{s \in
\mathbb{N}_0}$, $n\in\mathbb{N}_0$ is called {\it a hierarchical
structure} on $\mathbb{L}$. Given $l , l' \in \mathbb{L}$, we set
\begin{equation} \label{hd}
n (l, l') = \min\{n \ | \ \exists\Lambda_{n, s}: \ l, l' \in
\Lambda_{n,s}\}, \quad d(l,l') = \varkappa^{n(l,l')} - 1.
\end{equation}
The function $d:\mathbb{L}\times \mathbb{L} \rightarrow [0, +\infty)$ has the following property: any triple $\{l_1 ,
l_2 , l_3\}\subset \mathbb{L}$ contains two elements, say $l_1 , l_2$, such that $d(l_1 , l_3) = d(l_2 , l_3)$. Thus,
$d(l,l')$ is a metric on $\mathbb{L}$. The interaction potential in our model has the form of (\ref{ip}) with the above
metric $d(l,l')$. It is invariant under the transformations of $\mathbb{L}$ which leave $d(l,l')$ unchanged. In view of
this fact, it is convenient to choose the sequences $\mathcal{L}$ which determines the infinite-volume limit to be
consisting of the sets (\ref{h2}) only. A standard choice is the sequence of $\Lambda_{n , 0} \
\stackrel{\rm def}{=} \ \Lambda_n$, $n \in \mathbb{N}_0$.

The Hamiltonian (\ref{au1}) may be rewritten in the form
\begin{equation} \label{ip1}
H = -
\frac{\theta}{2}\sum_{n=0}^{\infty}\varkappa^{-n(1+\delta)}\sum_{l
\in \mathbb{L}}\left(\sum_{l' \in \Lambda_{n,l}}\mathfrak{q}_{l'}
\right)^2 + \sum_{l \in \mathbb{L}}
\left[\frac{1}{2\mathfrak{m}}\mathfrak{p}_l^2 + \frac{a}{2}
\mathfrak{q}_l^2 + b \mathfrak{q}_l^4\right] ,
\end{equation}
where $\theta = J (1- \varkappa^{-(1+\delta)})>0$. The local Hamiltonians indexed by $\Lambda_{n, l}$ are obtained from
the above one by the corresponding  truncation of the sums. For our purposes, it is convenient to write them
recursively
\begin{equation} \label{h3}
H_{\Lambda_{n,l}} \ \stackrel{\rm def}{=} \ H_{n, l} = -
\frac{\theta}{2}\varkappa^{-n(1+\delta)}\left(\sum_{s\in
\Lambda_{n, l}}\mathfrak{q}_s\right)^2 +  \sum_{s \in \Lambda_{1,
l}}H_{n - 1, s},
\end{equation}
where the one particle Hamiltonian is
\begin{equation} \label{h4}
H_{0, l} =  \frac{1}{2\mathfrak{m}}\mathfrak{p}_l^2 +
\frac{a}{2}\mathfrak{q}_l^2 + b \mathfrak{q}_l^4 .
\end{equation}
The canonical pair $\mathfrak{p}_l $, $\mathfrak{q}_l$, as well as the Hamiltonian $ H_{0, l}$, are defined in the
complex Hilbert space $\mathcal{H}_l= L^2 (\mathbb{R})$ as unbounded operators, which are essentially self-adjoint on
the dense domain $C_0^\infty (\mathbb{R})$. The Hamiltonian $H_{n, l}$, $n \in \mathbb{N}$ is defined similarly but in
the space $\mathcal{H}_{n, l} = L^2 (\mathbb{R}^{|\Lambda_{n, l}|})$.

The local Gibbs state in $\Lambda_{n,l}$ at a given temperature $\beta^{-1}>0$ is defined on $\mathfrak{C}_{n,j}$ --
the
 $C^*$-algebra of bounded operators on $\mathcal{H}_{n,
 l}$, as follows
\begin{equation} \label{h6}
\varrho_{\beta , \Lambda_{n, l}} (A) = \frac{{\rm
trace}\left(A\exp\left(- \beta H_{n, l} \right)\right)}{{\rm trace}\exp\left(- \beta H_{n, l} \right)}, \quad A\in
\mathfrak{C}_{n,l}.
\end{equation}
In a standard way, it may be extended to unbounded operators such as $\mathfrak{q}_{l'}$, $l' \in \Lambda_{n,l}$. The
dynamics in $\Lambda_{n, l}$ is described by the time automorphisms of $\mathfrak{C}_{n,l}$
\begin{equation} \label{h7}
\mathfrak{a}^t_{n, l} (A) = \exp\left( it H_{n, l} \right) A
\exp\left( - it H_{n, l} \right),\quad t\in \mathbb{R}.
\end{equation}
For a measurable function $A:\mathbb{R}^{|\Lambda_{n,l}|}\rightarrow \mathbb{C}$, the multiplication operator $A$ acts
on $\psi\in \mathcal{H}_{n, l}$ as $$ (A\psi)(x) = A(x) \psi(x), \ \quad x \in
\mathbb{R}^{|\Lambda_{n,l}|}. $$ It appears that the linear span
of the operators
\[
\mathfrak{a}^{t_1}_{n, l} (A_1)\cdots \mathfrak{a}^{t_k}_{n, l}
(A_k), \quad k \in \mathbb{N}, \ \ t_1 , \dots , t_k \in
\mathbb{R},
\]
with all possible choices of $k$, $t_1 , \dots t_k$ and multiplication operators  $A_1 , \dots , A_k \in
\mathfrak{C}_{n,l}$ is dense in the algebra $\mathfrak{C}_{n,l}$
in the $\sigma$-weak topology, in which the state (\ref{h6}) is continuous. Thus, this state is fully determined by
temporal Green functions
\begin{equation} \label{h8}
G_{A_1 , \dots , A_k}^{n, l}(t_1 , \dots , t_k ) =
\varrho_{\beta , \Lambda_{n, l}}( \mathfrak{a}^{t_1}_{n, l} (A_1)\dots
\mathfrak{a}^{t_k}_{n, l} (A_k)),
\end{equation}
corresponding to all possible multiplication operators $A_1 ,
\dots , A_k \in \mathfrak{C}_{n,l}$. Set
\begin{equation} \label{h9}
\mathcal{D}_k^\beta = \{(t_1 , \dots , t_k) \in \mathbb{C}^k\ | \
0< {\rm Im}(t_1) < \dots {\rm Im}(t_k)<\beta \}.
\end{equation}
As was proven in Lemma 2.1 in \cite{AKKR}, every Green function (\ref{h8}) may be extended to a holomorphic function on
$\mathcal{D}_k^\beta$. This extension is continuous on the closure of $\mathcal{D}_k^\beta$ and may be uniquely
determined by its values on the set
\begin{equation} \label{h10}
\mathcal{D}_k^\beta (0) = \{(t_1 , \dots , t_k)
\in \mathcal{D}_k^\beta \ | \ {\rm Re}(t_j) = 0 , \quad j = 1,
\dots , k \}.
\end{equation}
The restriction of $ G_{A_1 , \dots , A_k}^{n, l}$ to $\mathcal{D}_k^\beta (0 )$, i.e., the function
\begin{equation} \label{h10a}
\Gamma_{A_1 , \dots , A_k}^{n, l} (\tau_1 , \dots , \tau_k) =
G_{A_1 , \dots , A_k}^{n, l}(i\tau_1 , \dots , i\tau_k),
\end{equation}
is the Matsubara function corresponding to the operators $A_1, \dots , A_k$. By (\ref{h6}) - (\ref{h8}), it may be
written
\begin{eqnarray} \label{h10b}
& & \Gamma_{A_1 , \dots , A_k}^{n, l}(\tau_1, \dots , \tau_k) =
\frac{1}{Z_{n, l}}{\rm trace}\left\{A_1 \exp\left(- (\tau_2 -
\tau_1)H_{n, l} \right) \right.\\ & & \quad \left. \times
A_2\exp\left(- (\tau_3 - \tau_2)H_{n, l} \right)\dots A_k\exp\left(- (\beta - \tau_k + \tau_{1})H_{n, l}
\right)\right\}; \nonumber \\ & & \qquad \quad Z_{n, l} \
\stackrel{\rm def}{=} \ {\rm trace}\left\{\exp\left(- \beta H_{n,
l}\right)\right\}.\nonumber
\end{eqnarray}
This representation immediately yields the `KMS-periodicity'
\begin{equation} \label{h11}
\Gamma_{A_1 , \dots , A_k}^{n, l}(\tau_1 + \vartheta, \dots ,
\tau_k+ \vartheta) = \Gamma_{A_1 , \dots , A_k}^{n, l}(\tau_1,
\dots , \tau_k),
\end{equation}
for every $\vartheta\in \mathcal{I}_\beta \ \stackrel{\rm def}{=}\ [0, \beta] $, where addition is of modulo $\beta$.

As was mentioned in the introduction, the phase transition in the model is connected with the appearance of macroscopic
displacements of particles from their equilibrium positions $\mathfrak{q}_l =0$, which occur when the fluctuations of
such displacements become large. To describe them, we set (c.f., (\ref{au3}))
\begin{equation} \label{h12}
Q_{n, l}^\lambda = \frac{\lambda_n}{\sqrt{|\Lambda_{n,l}|}}
\sum_{l'\in \Lambda_{{n, l}}}\mathfrak{q}_{l'} =
\frac{\lambda_n}{\varkappa^{n/2}} \sum_{l'\in \Lambda_{{n,
l}}}\mathfrak{q}_{l'},
\end{equation}
where $\{\lambda_n\}_{n\in \mathbb{N}_0}$ is a sequence of positive numbers. The operators  $Q_{n, l}^\lambda$ are
unbounded, nevertheless, the corresponding Matsubara functions still possess almost all of those `nice' properties
which they have in the case of bounded operators. The next statement follows directly from Corollary 4.1 and Theorem
4.2 of \cite{AKKR}.
\begin{proposition} \label{unbound}
For every $n\in \mathbb{N}_0$ and $k\in \mathbb{N}$, the functions $\Gamma_{Q_{n, l}^\lambda, \dots ,Q_{n,
l}^\lambda}^{n, l}$ are continuous on $\mathcal{I}_\beta^{2k}$, they can be analytically continued to the domains
$\mathcal{D}^\beta_{2k}$.
\end{proposition}
The convergence of the sequence $\{\Gamma_{Q_{n, j}^\lambda, \dots ,Q_{n, j}^\lambda}^{n, j}\}_{n\in \mathbb{N}_0} $
with $\lambda_n = \varkappa^{-n/2}$, to a nonzero limit would mean the appearance of the long range order caused by
macroscopic displacements of particles. The convergence with a slower decaying sequence $\{\lambda_n\}$ corresponds to
the presence of a critical point.

Our model is described by the following parameters: $\delta>0$, which determines the decay of the potential $J_{ll'}$,
see (\ref{ip}); $\theta\geq 0$, which determines its strength; the mass $\mathfrak{m}$ and the parameters of the
potential energy ${a}$ and ${b}$, see e.g., (\ref{au1}). Since the choice of $\theta$ determines only the scale of
$\beta$, we may set
\begin{equation}
 \label{norm1} \theta =    \varkappa^{\delta} -1 ,
\end{equation}
which corresponds to the choice (see (\ref{au5}))
\[
J = J_* \ \stackrel{\rm def}{=} \
 \frac{ \varkappa^{\delta} -1 }{1 - \varkappa^{-1-\delta}}.
\]
To simplify notations we write the operator (\ref{h12}) with $\lambda_n = \varkappa^{-n\delta/2}$ as $Q_{n,l}$ and
\begin{equation} \label{k1}
\Gamma^{n,l}_{Q_{n,l}, \dots , Q_{n,l}} (\tau_1, \dots ,
\tau_{2k}) = \Gamma^{(n)}_{2k} (\tau_1, \dots , \tau_{2k}).
\end{equation}
\begin{theorem}\label{h1tm}
For the model (\ref{au1}) with $\delta\in (0, 1/2)$, one can choose the parameters  ${a}$, ${b}$ and $\mathfrak{m}$ in
such a way that there will exist $\beta_*>0$, dependent on $a$, $b$, $\mathfrak{m}$, with the following properties: (a)
if $\beta =
\beta_*$, then for all $k\in \mathbb{N}$, the convergence
\begin{equation} \label{h14}
\Gamma_{2k}^{(n)}(\tau_1 , \dots , \tau_{2k}) \longrightarrow
 \frac{(2k)!}{k!2^k\beta_*^k },
\end{equation}
holds uniformly on $(\tau_1 , \dots , \tau_{2k})\in
\mathcal{I}_\beta^{2k}$; (b) if $\beta < \beta_*$, the functions
$\Gamma_{2k}^{\alpha , \beta, \Lambda_{n, l}}$, $k \in \mathbb{N}$ defined by (\ref{au4}) converge to zero in the same
sense for all $\alpha
>0$.
\end{theorem}

\section{Euclidean Representation}

In the Euclidean approach \cite{AKKR} the functions (\ref{h10a}) corresponding to the multiplication operators $A_1 ,
\dots , A_{2k}$, are written as follows
\begin{equation} \label{h15}
\Gamma_{A_1, \dots ,A_{2k}}^{n, l}(\tau_1 , \dots , \tau_{2k}) =
\int_{\Omega_{n, l}}A_1 (\omega_{n, l} (\tau_1) )\dots A_{2k}
(\omega_{n, l} (\tau_{2k}) )\nu_{n, l} ({\rm d}\omega_{n, l}),
\end{equation}
where $\Omega_{n,l}$ is the Banach space of real valued continuous periodic functions
\begin{eqnarray} \label{cpf}
& & \Omega_{n,l}  =  \{ \omega_{n,l} = (\omega_{l'})_{l' \in
\Lambda_{n,l}} \ | \ \omega_{l'} \in \Omega\}, \\ & & \Omega  =
\{\omega\in C(\mathcal{I}_\beta \rightarrow \mathbb{R}) \ | \
\omega (0) = \omega (\beta)\}. \nonumber
\end{eqnarray}
The probability measure $\nu_{n,l}$ is
\begin{eqnarray} \label{h16}
\nu_{n, l} ({\rm d}\omega_{n, l})& = & \frac{1}{Z_{n, l}} \exp\left[-
E_{n, l} (\omega_{n, l})  \right]\chi_{n, l} ({\rm d}\omega_{n, l}),\\ Z_{n, l} & = & \int_{\Omega_{n,l}} \exp\left[-
E_{n, l} (\omega_{n, l})  \right]\chi_{n, l} ({\rm d}\omega_{n, l}). \nonumber
\end{eqnarray}
 The functions
$E_{n, l} : \Omega_{n, l} \rightarrow \mathbb{R}$ are (c.f., (\ref{h3}))
\begin{eqnarray} \label{h17}
E_{n, j} (\omega_{n, j}) & = & - \frac{1}{2}\theta
\varkappa^{-n(1+ \delta)}\int_0^\beta \left(\sum_{l' \in
\Lambda_{n ,l}}\omega_{l'} (\tau) \right)^2 {\rm d}\tau
 + \sum_{s \in
 \Lambda_{1,l}}E_{n-1, s} (\omega_{n-1, s}), \nonumber \\
E_{0, s} (\omega_{s}) & = & \int_{0}^{\beta} \left(\frac{a-1}{2} [\omega_s (\tau)]^2 + b[\omega_s (\tau)]^4 \right){\rm
d} \tau.
\end{eqnarray}
We consider $\omega_{n, l}$ as vectors
 $(\omega_{n-k,
s})_{s\in  \Lambda_{k,l}}$ with $k = 1, 2, \dots , n$  and write $\omega_s $ for $\omega_{0, s}$. The measure $\chi_{n,
l}$ is
\begin{eqnarray} \label{chi}
\chi_{n, l} ({\rm d}\omega_{n, l}) = \bigotimes_{s\in
 \Lambda_{n,l}}\chi ({\rm d}\omega_{s}).
\end{eqnarray}
where $\chi$ is a Gaussian measure on $\Omega_{0,s} = \Omega$. Let $\mathcal{E}$ be the real Hilbert space $L^2
(\mathcal{I}_\beta)$. Then the Banach space of continuous periodic functions $\Omega$ can be considered, up to
embedding, as a subset of $\mathcal{E}$. The following family
\begin{equation} \label{base} e_q (\tau) =
\left\{
\begin{array}{ll}
 \sqrt{\frac{2}{\beta}}\cos q \tau, \quad
& q
>0;  \\[.3cm]
 -\sqrt{\frac{2}{\beta}}\sin q \tau, \quad & q
<0; \\[.3cm] 1/\sqrt{\beta}, \quad & q=0.
\end{array}
\right.
\end{equation}
with $q$ varying in the set
\begin{equation}\label{h32}
 \mathcal{Q} = \{q \ | \ q= \frac{2\pi}{\beta}\kappa, \ \
\kappa \in \mathbb{Z} \},
\end{equation}
is a base of $\mathcal{E}$. Given $q\in \mathcal{Q}$, let $P_q$ be the orthonormal projection on $e_q$. We define
$\chi$ to be the Gaussian measure\footnote{For a topological space, `measure defined on the space' means that the
measure is defined on its Borel $\sigma$-algebra.} on $\mathcal{E}$ with zero mean and with the covariance operator
\begin{equation} \label{co}
S = \sum_{q\in\mathcal{Q}} \frac{1}{\mathfrak{m} q^2 + 1} P_q .
\end{equation}
One can show (see Lemma 2.2 of \cite{AKKR}) that the measure $\chi$ is concentrated on $\Omega$, i.e., $\chi(\Omega) =
1$. On the other hand, as follows from the Kuratowski theorem (see Theorem 3.9, page 21 of \cite{Pa}), the Borel
$\sigma$-algebras of subsets of $\Omega$, generated by its own topology and by the topology induced from the Hilbert
space $\mathcal{E}$, coincide. Hence, one can consider $\chi$ also as a measure on $\Omega$. As such one, it appears in
the representation (\ref{chi}).

The fluctuation operator $Q_{n,l}$, defined by (\ref{h12}) with $\lambda_n = \varkappa^{-n\delta/2}$ is a
multiplication operator by the function $Q_{n,l}: \mathbb{R}^{|\Lambda_{n,l}|}\rightarrow
\mathbb{R}$
\begin{equation} \label{h19}
Q_{n,l}(\xi_{n,l}) =
 \varkappa^{-n(1+\delta)/2}\sum_{s\in \Lambda_{n,l}} \xi_s  =
\varkappa^{-(1+\delta)/2}\sum_{s\in \Lambda_{1,l}} Q_{n-1, s}(\xi_{n-1, s}).
\end{equation}
The representation (\ref{h15}) and the properties of the measures $\nu_{n,l}$, $\chi_{n,l}$, $\chi$ (see Lemma 2.3 and
the whole section 2.2 of \cite{AKKR}) yield the following statement.
\begin{proposition} \label{10pn}
For every fixed $\beta>0,\ \tau_1 , \dots , \tau_{2k}\in
\mathcal{I}_\beta$, the Matsubara functions (\ref{k1})
continuously depend on $\mathfrak{m} >0$, $a\in \mathbb{R}$ and $b>0$.
\end{proposition}
\begin{proposition} \label{k1pn}
For all $n \in \mathbb{N}_0$ and $k \in \mathbb{N} $, the functions (\ref{k1}) obey the estimates
\begin{equation} \label{k2}
0 \leq \Gamma^{(n)}_{2k} (\tau_1, \dots , \tau_{2k}) \leq
\sum_{\sigma} \prod_{l=1}^{k} \Gamma_2^{(n)} \left(\tau_{\sigma
(2l-1) } ,\tau_{\sigma (2l)} \right),
\end{equation}
which hold for all $\tau_1, \dots , \tau_{2k} \in
\mathcal{I}_\beta$. Here the sum is taken over all possible
partitions of the set $\{ 1, \dots , 2k\}$ onto unordered pairs.
\end{proposition}
The estimates (\ref{k2}) were proven in \cite{AKKR} as Theorems 6.2 (positivity) and 6.4 (Gaussian upper bound).

Since to prove our theorem we need the Matsubara functions corresponding to the operators $Q_{n,l}$ only, we may
restrict our study to the measures describing distributions of $Q_{n,l}$ given by (\ref{h19}). For $n \in \mathbb{N}_0$
and a Borel subset $C\subset \Omega$, let
\[
B_C = \{ \omega_{n , l}\in \Omega_{n , l} \ | \
\varkappa^{-n(1+\delta)/2} \sum_{s \in \Lambda_{n , l}}\omega_s
\in C \},
\]
which is a Borel subset of $\Omega_{n , l}$. Then we set
\[
\mu_n (C) =  \nu_{n , l}(B_C),
\]
which defines a measure on $\Omega$. By (\ref{h16}), (\ref{h17}), the measures $ \mu_n $ obey
 the following recursion relation
\begin{eqnarray} \label{h20}
 {\mu}_{n} ({\rm d}{\omega}) & = &  \frac{1}{ {Z}_{n}}
\exp\left(\frac{\theta}{2}\|{\omega}\|^2_{\mathcal{E}} \right)
{\mu}_{n-1}^{\star \varkappa} (\varkappa^{(1+
 \delta)/2}{\rm d}{\omega}),  \\
\label{h21} {\mu}_0 ({\rm d}\omega)& = &\frac{1}{{Z}_{0}} \exp\left( -
E_{0,s} (\omega)\right)\chi({\rm d}\omega),
\end{eqnarray}
where $\|\cdot\|_\mathcal{E}$ is the norm in the Hilbert space $\mathcal{E}= L^2(\mathcal{I}_\beta )$, the function
$E_{0,s}$ is given by (\ref{h17}), ${Z}_{n}$, $n\in \mathbb{N}$ are normalizing constants and $\star $ stands for
convolution. For obvious reasons, we drop the labels $l$ and $s$. Like the measure $\chi$, all $\mu_n$, $n\in
\mathbb{N}_0$ can be considered either as measures on the Hilbert space $\mathcal{E}$ concentrated on its subset
$\Omega$, or as measures on the Banach space $\Omega$. We have
\begin{equation} \label{se1}
\Gamma^{(n)}_{2k} (\tau_1 , \dots , \tau_{2k}) = \int_{\Omega}
\omega(\tau_1 ) \cdots \omega (\tau_{2k})\mu_n ({\rm d}\omega),
\end{equation}
and the function (\ref{au6}) may be written in the form
\begin{equation} \label{xx2}
\varphi_n^{(\delta)} (x) \ \stackrel{\rm def}{=} \ \varphi_n (x) =
\int_{\mathcal{E}} \exp((x, \omega)_\mathcal{E}) {\mu}_n ({\rm
d}\omega) = \int_{\Omega} \exp((x, \omega)_\mathcal{E}) {\mu}_n ({\rm d}\omega), \quad x \in \mathcal{E},
\end{equation}
where $(\cdot, \cdot)_{\mathcal{E}}$ is the scalar product in $\mathcal{E}$. Expanding its logarithm into the series
(\ref{au7}) we obtain the Ursell functions (c.f., (\ref{k1}))
\begin{eqnarray} \label{uf}
U^{\delta, \beta, \Lambda_{n,s}}_{2k} (\tau_1 , \dots ,
\tau_{2k})\ \stackrel{\rm def}{ = } \ U^{({n})}_{2k} (\tau_1 ,
\dots , \tau_{2k}), \quad k\in\mathbb{N}.
\end{eqnarray}
Correspondingly, the numbers (\ref{au10}) obtained from these functions are denoted by $\mathcal{U}^{(n)}_{2k}$. Each
function $U^{({n})}_{2k}$ can be written as a polynomial of the Matsubara functions $\Gamma^{(n)}_{2s}$, $s = 1 , 2,
\dots , k$ and vice versa. In particular,
\begin{eqnarray} \label{h24}
U_2^{(n)} (\tau_1, \tau_2) & = & \Gamma_2^{(n)} (\tau_1, \tau_2), \\ U_4^{(n)} (\tau_1,\dots , \tau_4) & = &
\Gamma_4^{(n)} (\tau_1,\dots , \tau_4) - \Gamma_2^{(n)} (\tau_1, \tau_2)
\Gamma_2^{(n)} (\tau_3, \tau_4)\nonumber
\\ & - & \Gamma_2^{(n)} (\tau_1, \tau_3)\Gamma_2^{(n)} (\tau_2, \tau_4) -
\Gamma_2^{(n)} (\tau_1, \tau_4)\Gamma_2^{(n)} (\tau_2,
\tau_3).\nonumber
\end{eqnarray}
In view of (\ref{h11}), the Matsubara and Ursell functions depend only on the periodic distances between $\tau_j$,
i.e., on $|\tau_i - \tau_j|_\beta = \min\{ |\tau_i - \tau_j|, \beta - |\tau_i -
\tau_j|\}$.

The proof of Theorem \ref{h1tm} is based on inequalities for the Matsubara and Ursell functions, which we obtain by
means of the lattice approximation method. Its main idea is to construct sequences of probability measures,
concentrated on finite dimensional subspaces of $\Omega_{n,l}$, which converge to the Euclidean measures $\nu_{n,l}$ in
such a way that the integrals (\ref{h15}) are the limits of the corresponding integrals taken with such approximating
measures. Then the latter integrals are being rewritten as moments of Gibbs measures of classical ferromagnetic models,
for which one has a number of useful inequalities. In such a way, these inequalities are transferred to the Matsubara
and Ursell functions. A detailed description of this method is given in Section 5 of \cite{AKKR}. Here we provide a
short explanation of its main elements. Given $N = 2L$, $L\in
\mathbb{N}$, set
\begin{equation} \label{la1}
\lambda_q^{(N)} = \left\{\mathfrak{m}
\left(\frac{2N}{\beta}\right)^2 \left[
\sin\left(\frac{\beta}{2N}\right)q\right]^2 +1 \right\}^{-1},
\end{equation}
and
\begin{equation} \label{la2}
S_N = \sum_{q \in \mathcal{Q}_N} \lambda_q^{(N)}P_q , \quad
\mathcal{Q}_N = \{ q = \frac{2\pi}{\beta} \kappa \ | \ \kappa =
-(L-1), \dots , L\},\end{equation} where the projectors $P_q$ are the same as in (\ref{co}). Now let $\chi_N$ be the
Gaussian measure on $\mathcal{E}$ with the covariance operator $S_N$. Let also $\chi_{n,l}^{(N)}$ be defined by
(\ref{chi}) with $\chi_N$ instead of $\chi$. By means of $\chi_{n,l}^{(N)}$, we define $\nu_{n,l}^{(N)}$ via
(\ref{h16}). Then by Theorem 5.1 of
\cite{AKKR}, one has
\begin{equation} \label{la3}
\int_{\Omega_{n,l}} Q_{n,l} (\omega_{n,l} (\tau_1)) \cdots Q_{n,l}
(\omega_{n,l} (\tau_{2k}))\nu_{n,l}^{(N)}({\rm d}\omega_{n,l})
\longrightarrow \Gamma^{(n)}_{2k} (\tau_1 , \dots, \tau_{2k}),
\end{equation}
pointwise on $\mathcal{I}_\beta^{2k}$ as $N\rightarrow +\infty$. On the other hand, one can write
\begin{equation} \label{la4}
{\rm LHS (\ref{la3})} = C_{2k,N} \sum_{\ell_1 , \dots \ell_{2k} }
\langle S_{\ell_1} \cdots S_{\ell_{2k}} \rangle,
\end{equation}
where $C_{2k,N} >0$ is a constant and $\langle \cdot \rangle$ stands for the expectation with respect to the local
Gibbs measure on $\Xi_{n,l}^{(N)} \ \stackrel{\rm def}{=} \ \Lambda_{n,l} \times
\{1, 2, \dots , N\}$ of a ferromagnetic model with the one
dimensional $\phi^4$ single-spin distribution. This type of single-spin distribution is determined by our choice of the
potential energy in (\ref{au1}), whereas the ferromagneticity is due to the fact that $J>0$ (see (\ref{ip}) and due to
our choice of the numbers (\ref{la1}). The sum in (\ref{la4}) is taken over the vectors $\ell_j = (\ell_j^{(1)},
\ell_j^{(2)})$, $j = 1 ,
\dots , 2k$ as follows. Their first components run through
$\Lambda_{n,l}$ and the second components are fixed at certain values from the set $\{1, \dots, N\}$, determined by the
corresponding $\tau_j$. Furthermore, the above expectations $\langle \cdot \rangle$ can be approximated by expectations
with respect to the ferromagnetic Ising model (classical Ising approximation \cite{SG,Simon}). Then the functions
$\Gamma^{(n)}_{2k}$ and $U^{(n)}_{2k}$ obey the inequalities which the moments and semi-invariants of the ferromagnetic
Ising model do obey. In particular, we have the following.
\begin{lemma}
\label{5.7lm} For all $n\in \mathbb{N}_0$ and $k\in \mathbb{N}$,
the following estimates hold for all values of the arguments $\tau, \tau', \tau_1, \dots , \tau_{2k} \in
\mathcal{I}_\beta$,
\begin{equation} \label{5.32}
\int_{\mathcal{I}_\beta^2} U^{(n)}_4 (\tau , \tau , \tau_1 ,
\tau_2 ){\rm d}\tau_1
 {\rm d}\tau_2
\ \  \leq \ \ \int_{\mathcal{I}_\beta^2} U^{(n)}_4 (\tau , \tau' ,
\tau_1 , \tau_2 ){\rm d}\tau_1
 {\rm d}\tau_2 ;
\end{equation}
\begin{equation}\label{5.33}
(-1)^{k-1} U^{(n)}_{2k} (\tau_1 , \tau_2 , \dots , \tau_{2k} )
\geq 0 .
\end{equation}
\end{lemma}
\begin{proof}
 For classical models with unbounded spins and
polynomial anharmonicity of the Ellis-Monroe type (for $\phi^4$-models, in particular), (\ref{5.32}) was proved  in
\cite{AKK1}. For the Ising model, the sign rule (\ref{5.33}) was
proved in \cite{Schl}. $\square$
\end{proof}
\begin{lemma} \label{4lm}
For all $n,l\in \mathbb{N}_0$, the function
\begin{eqnarray} \label{la6}
f_n (z) & = & \int_{\Omega_{n,l}} \exp\left(z \int_0^\beta Q_{n,l}(\omega_{n,l} (\tau)){\rm d}\tau \right)
\nu_{n,l}({\rm d}\omega_{n,l}) \\
& = & \int_{\mathcal{E}} \exp\left(z \int_0^\beta \omega (\tau){\rm d}\tau \right) \mu_{n}({\rm d}\omega), \nonumber
\end{eqnarray}
can be analytically continued to an even entire function of order less than two, possessing  purely imaginary zeros.
\end{lemma}
\begin{proof}
For the function (\ref{la6}), one can construct the lattice approximation (c.f., (\ref{la4}))
\begin{equation} \label{laa1}
f_n^{(N)} (z) = \langle \exp\left(z \sum_{\ell \in \Xi^{(N)}_{n , l}}S_\ell \right)\rangle,
\end{equation}
which converges, as $N \rightarrow +\infty$, to $f_n (z)$ for every $z \in \mathbb{R}$. For such $f_n^{(N)}$, the
property stated is known as the generalized Lee-Yang theorem \cite{LS}. The functions $f_n^{(N)}$ are ridge (crested),
with the ridge being the real axis. For sequences of such functions, their pointwise convergence on the ridge implies
via the Vitali theorem (see e.g., Proposition VIII.19 in \cite{Simon}) the uniform convergence on compact subsets of
$\mathbb{C}$, which by the Hurwitz theorem (see e.g., \cite{Beren}) gives the desired property of $f_n$. $\square$
\end{proof}
Set
\begin{eqnarray} \label{h33}
\hat{u}_n (q) & = &  \int_{0}^{\beta} U_2^{(n)} (\tau',
\tau) \cos (q \tau) {\rm d} \tau, \\ & = &
\int_{0}^{\beta} U_2^{(n)} (0, \tau) \cos (q \tau) {\rm d}
\tau , \quad q\in \mathcal{Q}, \quad n
\in
\mathbb{N}_0     . \nonumber
\end{eqnarray}
Then
\begin{equation} \label{h35}
U_2^{(n)} (\tau_1, \tau_2) = \frac{1}{\beta}\sum_{q \in
\mathcal{Q}}\hat{u}_n (q) \cos [q (\tau_1 - \tau_2)].
\end{equation}
Furthermore, we set (c.f., (\ref{au10}))
\begin{equation} \label{laa2}
\mathcal{U}^{(n)}_{2k} = \int_{\mathcal{I}_\beta^{2k}}
U^{(n)}_{2k} (\tau_1 , \dots , \tau_{2k}) {\rm d}\tau_1 \cdots {\rm d}\tau_{2k}.
\end{equation}
Then
\begin{equation} \label{kokos}
\hat{u}_n (0) = \hat{u}_n  \ \stackrel{\rm def}{=} \
\beta^{-1}\mathcal{U}_2^{(n)} = \beta^{-1}\mathcal{U}_2^{\delta,
\beta, \Lambda_n}.
\end{equation}
\begin{lemma}
 \label{ormlm}
 For every $n\in \mathbb{N}_0$ and $q\in \mathcal{Q}$, $\hat{u}_n (q)$ is
 a continuous function of $\beta$, it obeys the following
 estimates
\begin{eqnarray} \label{6.120}
& & 0 < \hat{u}_n (q) \leq \hat{u}_n ; \\
\label{20.331}
  & & \hat{u}_n (q) \leq
 \frac{\varkappa^{-n\delta}}{\mathfrak{m}q^2 }, \quad  q \neq 0.
\end{eqnarray}
\end{lemma}
\begin{proof} By
 (\ref{h33}), (\ref{h24}), (\ref{k1}) and (\ref{h10b}), one obtains
\begin{eqnarray*}
U_2^{(n)} (0, \tau )  = \frac{1}{Z_{n,l}}{\rm trace}\left\{Q_{n,l}
\exp\left[-\tau H_{n,l}\right] Q_{n,l} \exp\left[-(\beta -\tau)
H_{n,l}\right]\right\}. \nonumber
\end{eqnarray*}
It may be shown that every $H_{n,l} $ has a pure point spectrum $\{ E_p^{(n)}\}_{ p\in \mathbb{N}_0}$. We denote the
corresponding eigenfunctions by $\Psi_p^ {(n)}$ and set
$$ Q_{pp'}^{(n)} {=}  (Q_{n,l}\Psi_p^{(n)} ,
\Psi_{p'}^{(n)})_{\mathcal{H}_{n,l}} . $$ Then the above
representation may be rewritten $$ U_2^{(n)} (0,\tau) =
\frac{1}{Z_{n,l}} \sum_{p, p'\in \mathbb{N}_0}\left\vert
Q_{pp'}^{(n)} \right\vert^2 \exp\left[ -\beta E_{p}^{(n)} + \tau (E_{p}^{(n)} - E_{p'}^{(n)}) \right] ,
$$ which yields via (\ref{h33})
\begin{eqnarray} \label{ag3}
\hat{u}_n (q)& = &\frac{1}{Z_{n,l}} \sum_{p,p'\in
\mathbb{N}_0}\left\vert Q_{pp'}^{(n)} \right\vert^2
\frac{E_p^{(n)} - E_{p'}^{(n)}}{q^2 + (E_p^{(n)} - E_{p'}^{(n)})^2
}
\\ & & \quad \times  \left(\exp[-\beta E_{p'}^{(n)}] - \exp[-\beta  E_{p}^{(n)
}]\right), \nonumber\\ Z_{n,l} & = & \sum_{p\in
\mathbb{N}_0}\exp[-\beta E_{p}^{(n) }]. \nonumber
\end{eqnarray}
Both series above converge uniformly, as functions of $\beta$, on compact subsets of $(0, +\infty)$, which yields
continuity and positivity. The upper bound (\ref{6.120}) follows from (\ref{ag3}) or from (\ref{h33}). To prove
(\ref{20.331}), we estimate the denominator in (\ref{ag3}) from below by $q^2 \neq 0$ and obtain
\begin{eqnarray}
\label{oldx} \hat{u}_n (q) &\leq & \frac{1}{q^2} \frac{1}{Z_{n,l}}
\sum_{p,p'} \left\vert Q_{pp'}^{(n)}\right\vert^2 (E_p^{(n)} -
E_{p'}^{(n)}) \left(\exp[-\beta E_{p'}^{(n)}] - \exp[-\beta E_{p}^{(n)} ]\right) \nonumber \\ &=&  \frac{1}{q^2}
\frac{1}{Z_{n,l}}{\rm trace}\left\{\left[Q_{n,l}, \left[H_{n,l},
Q_{n,l}\right]\right]\exp\left(-\beta H_{n,l}\right)\right\},
\quad q \neq 0.
\end{eqnarray}
By means of (\ref{h3}) and (\ref{au2}), the double commutator in (\ref{oldx}) may be computed explicitly. It equals to
$\left|
\Lambda_{n, l} \right|^{-\delta }/\mathfrak{m} $, which yields
(\ref{20.331}).
 $\square$
\end{proof}
\begin{lemma} \label{5.10lm}
The numbers $\mathcal{U}^{(n)}_{2k}$ defined by (\ref{laa2}) obey the estimates (c.f., (\ref{au14z}))
\begin{eqnarray} \label{5.351}
|\mathcal{U}^{(n)}_{2k} | & \leq & 2^{1-k}(2k-1)! (\beta \hat{u}_n )^{k}, \quad k\in \mathbb{N},\\
\label{la10}
 |\mathcal{U}^{(n)}_{2k} | & \leq&
\ \frac{ (2k-1)!}{ 3\cdot 2^{k-1}} (\beta \hat{u}_n )^{k-2}
|\mathcal{U}^{(n)}_{4} | , \quad k\geq 2.
\end{eqnarray}
\end{lemma}
\begin{proof}
The function (\ref{la6}) is the same as in (\ref{au11}), hence, it possesses the representation (\ref{au12}) and
$\mathcal{U}^{(n)}_{2k}= \mathcal{U}^{\delta, \beta,
\Lambda_{n,l}}_{2k}$ are given by the right-hand side of
(\ref{au13}). Then the estimates (\ref{5.351}), (\ref{la10}) immediately follow from (\ref{au14z}). $\square$
\end{proof}

\section{Proof of the Theorem}

Set
\begin{equation}\label{h36a}
X_n = - \int_{\mathcal{I}^2_\beta} U_4^{(n)} (\tau , \tau , \tau_1 , \tau_2){\rm d} \tau_1 {\rm d} \tau_2 .
\end{equation}
Then by Lemma \ref{5.7lm}, one has
\begin{equation} \label{au34}
0< \beta^{-2} \left\vert\mathcal{U}_4^{(n)} \right\vert \leq X_n,
\quad {\rm for \ all \ } n \in \mathbb{N}_0 ,
\end{equation}
thus, we may control the sequence $\{\mathcal{U}_4^{(n)}\}_{n \in
\mathbb{N}_0}$ by controlling $\{X_n \}_{n \in \mathbb{N}_0}$.
\begin{lemma}
\label{5.5lm} For the model (\ref{au1}) with $\delta\in (0, 1/2)$,
one can choose the parameters  ${a}$, ${b}$ and $\mathfrak{m}$ in such a way that there will exist $\beta_*>0$,
dependent on $a$, $b$, $\mathfrak{m}$,  with the following properties:
 (a) for $\beta \leq \beta_{*}$,
$\{ X_n \}_{ n\in \mathbb{N}_0} \rightarrow 0$; (b) for $\beta =
\beta_*$, $\{\hat{u}_{n}\}_{ n\in \mathbb{N}_0} \rightarrow 1$;
for $\beta < \beta_*$, there exists $K(\beta) >0$ such that for all $n\in \mathbb{N}_0$,
\begin{equation} \label{neww}
\hat{u}_n  \leq K(\beta )\varkappa^{-n\delta } .\end{equation}
\end{lemma}
The proof of this lemma will be given in the concluding section of the article. Lemmas \ref{ormlm} and \ref{5.5lm} have
two important corollaries.
\begin{corollary} \label{k1co}
For every $\beta \leq \beta_*$ and $k \in \mathbb{N}$, the sequences $\{\Gamma_{2k}^{(n)} \}_{n \in \mathbb{N}_0}$,
$\{U_{2k}^{(n)} \}_{n \in \mathbb{N}_0}$ are relatively compact in the topology of uniform convergence on
$\mathcal{I}_\beta^{2k}$.
\end{corollary}
\begin{proof}
Since the Ursell functions $U^{(n)}_{2k}$ may be expressed as polynomials of $\Gamma_{2s}^{(n)}$ with $s = 1 , \dots ,
k$ and vice versa, it is enough to prove this statement for the Matsubara functions only. By Ascoli's theorem (see
e.g., \cite{Mujica} p. 72) we have to show that the sequence $\{\Gamma_{2k}^{(n)} \}_{n
\in \mathbb{N}_0}$ is pointwise bounded and equicontinuous. By
(\ref{20.331}) and (\ref{h35}),
\begin{equation} \label{k33}
\Gamma_2^{(n)} (\tau , \tau')\leq   \Gamma_2^{(n)} (0 , 0)  \leq
\frac{1}{\beta}\hat{u}_{n}  +
\frac{\varkappa^{-n\delta}}{\beta\mathfrak{m}}\sum_{q \in
\mathcal{Q} \setminus \{0\} }\frac{1}{q^2}.
\end{equation}
For $\beta\leq \beta_*$, the sequence $\{\hat{u}_{n}  \}_{n \in
\mathbb{N}_0}$ is bounded by Lemma \ref{5.5lm}. Together with the
Gaussian upper bound (\ref{k2}) this yields the uniform boundedness of $\Gamma_{2k}^{(n)}$ on $\mathcal{I}_\beta^{2k}
$. Further, by (\ref{se1})
\begin{eqnarray} \label{k4}
& & \Gamma_{2k}^{(n)} (\tau_1 , \dots, \tau_{2k}) -
\Gamma_{2k}^{(n)} (\vartheta_1 , \dots, \vartheta_{2k}) \\ & & =
\int_{\mathcal{E}}\sum_{l=1}^{2k} \omega (\tau_1 ) \cdots \omega
(\tau_{l-1} )\left[\omega (\tau_l ) - \omega (\vartheta_l )
\right]\omega (\vartheta_{l+1} ) \cdots \omega (\vartheta_{2k} )
{\mu}_n ({\rm d}\omega). \nonumber
\end{eqnarray}
Applying here the Schwarz inequality (as to the scalar product  in $L^2 (\mathcal{E},{\mu}_n) $ of $\left[\omega
(\tau_l ) - \omega (\vartheta_l )   \right]$ and the rest of $\omega$), the Gaussian upper bound (\ref{k2}) and the
left-hand inequality in (\ref{k33}) one gets
\begin{eqnarray} \label{k5}
& & |\Gamma_{2k}^{(n)} (\tau_1 , \dots, \tau_{2k}) -
\Gamma_{2k}^{(n)} (\vartheta_1 , \dots, \vartheta_{2k})|^2  \\ & &
\quad \leq  \left(\Gamma_{2}^{(n)} (0 , 0) -\Gamma_{2}^{(n)}
(\tau ,  \vartheta ) \right) \cdot \frac{8k^2(4k - 2)!}{(2k-1)! 2^{2k-1}}\left( \Gamma_2^{(n)}(0,0)\right)^{2k-1},
\nonumber
\end{eqnarray}
where $(\tau ,  \vartheta)$ is chosen amongst the pairs $(\tau_l ,
\vartheta_l)$, $l = 1,  \dots, 2k $ to obey $|\tau -
\vartheta|_\beta = \max_{l}|\tau_l - \vartheta_l|_\beta$. But by
(\ref{h35}), (\ref{20.331}),
\begin{eqnarray*}
\Gamma_{2}^{(n)} (0 , 0) -\Gamma_{2}^{(n)} (\tau ,  \vartheta )
 & = &  \frac{2}{\beta} \sum_{q \in \mathcal{Q}}\hat{u}_{n} (q) \left\{\sin\left[
(q/2)\left(\tau - \vartheta \right) \right]\right\}^2 \\ & \leq &
2\frac{\varkappa^{-n\delta}}{\beta\mathfrak{m}}\sum_{q \in
\mathcal{Q} \setminus \{0\} }\frac{1}{q^2}\left\{\sin\left[(
q/2)\left(\tau - \vartheta \right) \right]\right\}^2 \nonumber \\ & \leq & C \varkappa^{-n\delta} |\tau -
\vartheta|_\beta,
\end{eqnarray*}
with an appropriate $C>0$. $\square$
\end{proof}
\vskip.1cm The next fact follows immediately from (\ref{20.331})
and (\ref{h32}).
\begin{corollary}
\label{N2co} For every $\beta$, $$ \sum_{q\in \mathcal{Q}\setminus
\{0\}} \hat{u}_n (q)
 \longrightarrow  0 , \ \ \ n\rightarrow +\infty .
$$
\end{corollary}
\vskip.1cm {\it Proof of Theorem \ref{h1tm}.} By Lemma
\ref{5.5lm}, (\ref{la10}), and (\ref{h36a}), (\ref{au34}), one
obtains that for all $k\geq 2$ and $\beta\leq \beta_* $, $\{\mathcal{U}_{2k}^{(n)}\}_{n\in\mathbb{N}_0} \rightarrow 0$.
Then by the sign rule (\ref{5.32}), for all $k \geq 2$, the sequences $\{{U}_{2k}^{(n)}\}_{n\in\mathbb{N}_0}$ converge
to zero for almost all $(\tau_1 , \dots , \tau_{2k})\in
\mathcal{I}_\beta^{2k}$, which, by Corollary \ref{k1co}, yields
their uniform convergence to zero. By (\ref{h35}) -- (\ref{20.331}), Corollary \ref{N2co}  and Lemma \ref{5.5lm}, one
has for $\beta = \beta_* $,
\begin{equation} \label{k7}
U_2^{(n)} (\tau_1 , \tau_2) = \frac{1}{\beta}\hat{u}_n  +
\frac{1}{\beta}\sum_{q\in \mathcal{Q}\setminus  \{0\}}\hat{u}_n
(q)\cos[q(\tau_1 - \tau_2 )] \longrightarrow  1/ \beta_*,
\end{equation}
uniformly on $\mathcal{I}_\beta^2$. Now one can express each $\Gamma_{2k}^{(n)}$ polynomially by $U_{2l}^{(n)}$ with $l
= 1 ,
\dots , k$ and obtain the convergence (\ref{h14}) for $\beta =
\beta_*$. For $\beta < \beta_*$, we have the estimate
(\ref{neww}), which yields (c.f., (\ref{k33}))
\begin{equation} \label{k700}
\Gamma_2^{\alpha, \beta, \Lambda_{n,l}} (\tau , \tau')\leq
\Gamma_2^{\alpha, \beta, \Lambda_{n,l}}  (0 , 0) \leq
\frac{\varkappa^{- n \alpha}}{\beta} \left[ K(\beta)  +
\frac{1}{\mathfrak{m}}\sum_{q \in \mathcal{Q} \setminus \{0\}
}\frac{1}{q^2}\right],
\end{equation}
hence $\Gamma_2^{\alpha, \beta, \Lambda_{n,l}} (\tau , \tau')
\rightarrow 0$ as $n \rightarrow +\infty$, uniformly on
$\mathcal{I}_\beta^2$. The convergence of the Matsubara functions $ \Gamma_{2k}^{\alpha, \beta, \Lambda_{n,l}}$ with
$k\geq 2$ follows from the Gaussian upper bound (\ref{k2}). $\square$

\section{Proof of Lemma \ref{5.5lm}}
Set
\begin{equation} \label{h28}
\sigma(v) = \frac{\varkappa^{-\delta}}{1 - ( 1 -
\varkappa^{-\delta})v}, \quad v \in \left( 0, ( 1 -
\varkappa^{-\delta})^{-1}\right),
\end{equation}
and
\begin{eqnarray} \label{au30}
\phi (v) =  \varkappa^{2 \delta -1}\left[\sigma (v) \right]^4,
\quad \psi(v)  =  \frac{1}{2} \varkappa^{2\delta -1}(1 -
\varkappa^{-\delta}) \left[\sigma (v) \right]^3.
\end{eqnarray}
\begin{lemma}
\label{6.2lm} Given $n\in \mathbb{N} $, let the condition
\begin{equation} \label{ag1}
\hat{u}_{n-1} (1-\varkappa^{-\delta})<1,
\end{equation}
be satisfied. Then the following inequalities hold:
\begin{equation} \label{6.8}
\hat{u}_n < \sigma \left(\hat{u}_{n-1} \right)
\hat{u}_{n-1};\end{equation}
\begin{equation} \label{6.10} \hat{u}_n  \geq
\sigma\left(\hat{u}_{n-1}\right) \hat{u}_{n-1} - \psi
(\hat{u}_{n-1} ) X_{n-1};\end{equation}
\begin{equation} \label{6.11} 0< X_n \leq \phi (\hat{u}_{n-1} )
X_{n-1}  ;
\end{equation}
where $\sigma(v)$, $\psi (v)$, $\phi (v)$ and $X_n$ are defined by (\ref{h28}), (\ref{au30}) and (\ref{h36a})
respectively.
\end{lemma}
\begin{proof}
For $t\in [0,\theta ]$, $\theta = \varkappa^\delta -1$, $x \in
\mathcal{E}$ and $n\in \mathbb{N}$, we set (c.f., (\ref{xx2}))
\begin{equation} \label{12}
\varphi_n (x |t) = \frac{1}{Z_n} \int_{\mathcal{E}} \exp\left( (x,
\omega)_\mathcal{E} + \frac{t}{2} \|\omega\|_\mathcal{E}^2
\right)\mu_{n-1}^{\star
\varkappa}\left(\varkappa^{(1+\delta)/2}{\rm d}\omega) \right),
\end{equation}
where $Z_n$ is the same as in (\ref{h20}). Then
\begin{equation} \label{13} \varphi_n (x|\theta) = \varphi_n (x),
\quad \varphi_n (x|0)=  Z_n^{-1} \left[\varphi_{n-1} \left(
\varkappa^{- (1+\delta)/2} x \right)\right]^\varkappa.
\end{equation}
For every $t\in [0,\theta]$, the function (\ref{12}) can be expanded in the series (\ref{au6}) with the coefficients
\begin{eqnarray} \label{14}
\varphi^{(n)}_{2k} (\tau_1 , \dots , \tau_{2k} |t) & = &
\frac{1}{Z_n}  \int_{\mathcal{E}}\omega (\tau_1 ) \cdots
\omega(\tau_{2k}) \exp\left(  \frac{t}{2} \|\omega\|_\mathcal{E}^2
\right) \nonumber\\ & \times &\mu_{n-1}^{\star
\varkappa}\left(\varkappa^{(1+\delta)/2}{\rm d}\omega) \right),
\end{eqnarray}
which, by (\ref{se1}), coincide with the corresponding Matsubara functions for $t = \theta$. For every fixed $(\tau_1 ,
\dots ,
\tau_{2k}) \in \mathcal{I}_\beta^{2k}$, as functions of $t$ they
are differentiable at any $t\in (0, \theta)$ and continuous on $[0, \theta]$. The corresponding derivatives are
obtained from (\ref{14})
\begin{eqnarray} \label{15}
& & \frac{\partial}{\partial t} \varphi^{(n)}_{2k} (\tau_1 , \dots , \tau_{2k} |t)
\ \stackrel{\rm def}{=}
\ \dot{\varphi}^{(n)}_{2k} (\tau_1 , \dots , \tau_{2k} |t) \\
& & \qquad \qquad = \frac{1}{2} \int_0^\beta \varphi^{(n)}_{2k+2} (\tau_1 , \dots , \tau_{2k}, \tau , \tau |t){\rm
d}\tau .
\nonumber
\end{eqnarray}
Now we write $\log \varphi_n (x |t)$ in the form of the series (\ref{au7}) and obtain the Ursell function $u_{2k}^{(n)}
(\tau_1 ,
\dots , \tau_{2k} |t)$. The derivatives of these functions with
respect to $t$ are being calculated from (\ref{15}). In particular, this yields
\begin{eqnarray} \label{lev5}
& & \dot{u}_{2}^{(n)} ( \tau_1,  \tau_{2} |t)= \\ & &  \quad = \frac{1}{2}\int_{0}^{\beta} {u}_{4}^{(n)} ( \tau_1,
\tau_{2}, \tau, \tau |t){\rm d}\tau + \int_{0}^\beta{u}_{2}^{(n)}
( \tau_1, \tau |t) {u}_{2}^{(n)} ( \tau_2,  \tau |t){\rm d}\tau ;
\nonumber
\end{eqnarray}
\begin{eqnarray} \label{lev6}
& & \dot{u}_{4}^{(n)} ( \tau_1, \tau_2, \tau_3,  \tau_{4} |t) = \\
 & & \quad = \frac{1}{2} \int_{0}^{\beta} u_{6}^{(n)} ( \tau_1,
\tau_2, \tau_3,  \tau_{4}, \tau , \tau |t){\rm d}\tau +  \nonumber \\
& & \quad + \int_{0}^{\beta} u_{4}^{(n)} ( \tau_1, \tau_2, \tau_3,
\tau |t ) {u}_{2}^{(n)} ( \tau_4,  \tau |t){\rm d}\tau   +
\nonumber \\& & \quad +\int_{0}^{\beta} u_{4}^{(n)} ( \tau_1,
\tau_2, \tau_4, \tau |t ) {u}_{2}^{(n)} ( \tau_3, \tau |t){\rm
d}\tau +  \nonumber \\ & & \quad +\int_{0}^{\beta} u_{4}^{(n)} (
\tau_1, \tau_3, \tau_4,
\tau |t ) {u}_{2}^{(n)} (\tau_2, \tau|t){\rm d}\tau +  \nonumber \\
& & \quad  + \int_{0}^{\beta} u_{4}^{(n)} ( \tau_2, \tau_3,
\tau_4, \tau |t ) {u}_{2}^{(n)} ( \tau_1, \tau |t){\rm d}\tau .
\nonumber
\end{eqnarray}
Then for
\begin{equation} \label{lev7}
{\upsilon}_n (t) \ \stackrel{\rm def}{ =} \
\int_{0}^{\beta}{u}_{2}^{(n)} ( \tau_1, \tau_2|t){\rm d}\tau_2 =
\int_{0}^{\beta}{u}_{2}^{(n)} ( 0, \tau |t){\rm d}\tau,
\end{equation}
we obtain the following system of equations
\begin{eqnarray} \label{lev8}
& & \dot{\upsilon}_n (t)  = \frac{1}{2} U(t) +  \left[\upsilon_n (t)\right]^2 ,\\ \label{lev9} & & \dot{U}(t) =
\frac{1}{2} V (t) + 2\upsilon_n (t)U(t) + \\ & & \qquad  + 2\int_{\mathcal{I}_\beta^{3}} u_2^{(n)} (\tau_2 ,
\tau_3|t)u_{4}^{(n)} ( 0, \tau_1, \tau_2, \tau_3 |t ){\rm d}\tau_1
{\rm d}\tau_2 {\rm d}\tau_3 , \nonumber
\end{eqnarray}
subject to the initial conditions (see (\ref{13}))
\begin{eqnarray} \label{lev13}
& & \upsilon_n (0) = \varkappa^{-\delta} \hat{u}_{n-1} , \\ & & U(0) =
\varkappa^{-2\delta-1}\int_{0}^{\beta}u_{4}^{(n)} ( 0,
\tau_1, \tau_2, \tau_2 |t){\rm d}\tau_1 {\rm d}\tau_2 = -
\varkappa^{-2\delta-1}X_{n-1}. \nonumber
\end{eqnarray}
Here
\begin{eqnarray} \label{lev10}
U(t)& \stackrel{\rm def}{=} &
\int_{\mathcal{I}_\beta^{2}}u_{4}^{(n)} ( 0, \tau_1, \tau_2,
\tau_2 |t ){\rm d}\tau_1 {\rm d}\tau_2 =  \\ &  & \quad =
\int_{\mathcal{I}_\beta^{2}}u_{4}^{(n)} ( \tau , \tau ,
\tau_1,
\tau_2 |t){\rm d}\tau_1 {\rm d}\tau_2 ,  \nonumber \\
 V (t)& \stackrel{\rm def}{=}& \int_{\mathcal{I}_\beta^{3}}u_{6}^{(n)} ( 0, \tau_1,
\tau_2, \tau_2,  \tau_{3}, \tau_3 |t ){\rm d}\tau_1 {\rm
d}\tau_2{\rm d}\tau_3 . \nonumber
\end{eqnarray}
Along with the problem (\ref{lev8}), (\ref{lev13}) we consider the following one
\begin{equation} \label{lev12}
\dot{y}(t) = [y(t)]^2, \quad y(0) = \upsilon_{n} (0) =
\varkappa^{-\delta}\hat{u}_{n-1}  .
\end{equation}
Under the condition (\ref{ag1}) it has a solution
\begin{equation} \label{ag2}
y(t) =
\frac{\varkappa^{-\delta}\hat{u}_{n-1}}{1-t\varkappa^{-\delta}\hat{u}_{n-1}
} = \sigma ((t/\theta) \hat{u}_{n-1}) \hat{u}_{n-1}, \quad t \in [0, \theta].
\end{equation}
The sign rule (\ref{5.33}) is valid for the above $u_{2k}^{(n)}$ for all $t\in [0, \theta]$, which yields $U(t) < 0$,
$V(t) > 0 $. Therefore, the solution of (\ref{lev8}) will be dominated\footnote{A detailed presentation of methods
based on differential inequalities are given in \cite{Wa}.} by (\ref{ag2}), i.e.,
\[
 \hat{u}_{n}  = \upsilon_{n} (\theta) < y(\theta) = \sigma
(\hat{u}_{n-1} )\hat{u}_{n-1} ,
\]
that gives (\ref{6.8}). Further, with the help of (\ref{5.32}), (\ref{5.33}) the third term on the right-hand side of
(\ref{lev9}) may be estimated as follows
\begin{eqnarray*}
& & 2 \int_{\mathcal{I}_\beta^{2}}u_{2}^{(n)} ( \tau_2, \tau_3 |t)\left(
\beta^{-1}\int_{\mathcal{I}_\beta^{2}}u_{4}^{(n)} (
\tau , \tau_1, \tau_2,  \tau_{3}|t){\rm d}\tau {\rm d}\tau_1
\right){\rm d}\tau_2{\rm d}\tau_3 \geq \\ & & \quad  \geq 2 \left(
\beta^{-1}\int_{\mathcal{I}_\beta^{2}}u_{4}^{(n)} (  \tau ,
\tau_1, \tau_2,  \tau_{2}|t){\rm d}\tau {\rm d}\tau_1 \right)
\times
\\ & & \quad \times \int_{\mathcal{I}_\beta^{2}}u_{2}^{(n)} ( \tau_2, \tau_3|t){\rm
d}\tau_2{\rm d}\tau_3= 2 \upsilon_n (t)U(t).
\end{eqnarray*}
Applying this in (\ref{lev9}) we arrive at (recall that $U(t)< 0$ and $V(t) > 0$)
\begin{equation}
\label{hhh}
 \frac{\dot{U}(t)}{U(t)} \leq 4 y(t) =  \frac{4 \varkappa^{-\delta}
\hat{u}_{n-1}}{1 - t \varkappa^{-\delta} \hat{u}_{n-1}}, \quad
\forall  t \in [0, \theta].
\end{equation}
Integrating one gets
\begin{equation} \label{tan}
U(t) \geq \frac{U(0)}{[1 -t \varkappa^{-\delta} \hat{u}_{n-1}]^4 }, \quad \forall t \in [0, \theta],
\end{equation}
which yields in turn
\begin{eqnarray*}
U(\theta)  =  - X_n \geq - \varkappa^{2\delta -1}\left[\sigma(\hat{u}_{n-1} ) \right]^4 X_{n-1} = -
\phi(\hat{u}_{n-1} ) X_{n-1},
\end{eqnarray*}
that gives (\ref{6.11}). Now we set
\[
 h(t) = \frac{1}{[1+t \varkappa^{-\delta}\hat{u}_{n-1}]^2}\upsilon_n
 \left( \frac{t}{1+t \varkappa^{-\delta}\hat{u}_{n-1}}\right) -
\frac{\varkappa^{-\delta}\hat{u}_{n-1}}{1+t
\varkappa^{-\delta}\hat{u}_{n-1}},
\]
where $t \in [0,t_{\rm max}]$, $t_{\rm max}= \theta
\varkappa^\delta \sigma (\hat{u}_{n-1})$. For this function, we
obtain from (\ref{lev8}) the following equation
\begin{equation}
\label{hhh1}
 \dot{h} (t) = \frac{1}{2 [1+t \varkappa^{-\delta}\hat{u}_{n-1}]^4}
 U\left(\frac{t}{1+t \varkappa^{-\delta}\hat{u}_{n-1}}\right) + [h(t)]^2,
\end{equation}
subject to the boundary conditions
\begin{equation} \label{hhh2}
h(0) = 0, \quad h(t_{\rm max}) = \left[1-\theta
\varkappa^{-\delta}\hat{u}_{n-1} \right]^2 \left[\upsilon_n
(\theta ) - \sigma(\hat{u}_{n-1})\hat{u}_{n-1} \right].
\end{equation}
By means of (\ref{hhh}), one may show that the first term on the right-hand side of (\ref{hhh1}) is a monotone
increasing function of $t\in [0, t_{\rm max}]$, which yields
\[
h(t_{\rm max}) - h(0) \geq  t_{\rm max} U(0)/2.
\]
Taking into account (\ref{hhh2}) and (\ref{lev13}) one obtains from the latter
\begin{eqnarray*}
 \upsilon_n (\theta) - \sigma(\hat{u}_{n-1})\hat{u}_{n-1}
& = &\hat{u}_n  - \sigma(\hat{u}_{n-1})\hat{u}_{n-1} \geq
\\ &\geq & - \frac{1}{2} (1-\varkappa^{-\delta})[\sigma(\hat{u}_{n-1})]^3
\varkappa^{2\delta -1} X_{n-1},
\end{eqnarray*}
that gives (\ref{6.10}). $\square$
\end{proof}
Now we prove a statement, which will allow us to control the initial elements in the sequences $\{\hat{u}_n \}$,
$\{X_n\}$, i.e., $\hat{u}_0$ and $X_0$. Set
\begin{equation} \label{1}
\eta = \eta (\beta, \mathfrak{ m}, a, b) = \varrho_{\beta ,
\Lambda_{0,l}}(\mathfrak{q}_{l}^2) = \int_{\Omega} [\omega
(0)]^{2}\mu_0 ({\rm d} \omega).
\end{equation}
From now on we suppose that $a<0$. Set also
\begin{equation} \label{3}
f(t) = t^{-1} \left(1 - e^{-t} \right) .
\end{equation}
\begin{lemma} \label{alm}
The following estimates hold
\begin{equation} \label{4}
\frac{\beta |a|}{12 b} f\left(\frac{3 \beta b}{\mathfrak{m}|a|
}\right)\leq \hat{u}_0 \leq \min\left\{ \beta \eta ; \
\frac{\beta|a| }{8 b}\left[1 + \sqrt{1 + (16 b/ \beta |a|)}\right]
\right\},
\end{equation}
\begin{equation} \label{5}
X_0 \leq 4! b\hat{u}_0^4 \left[ f \left(\frac{3 \beta b}{
\mathfrak{m}|a |}\right)\right]^{-1}.
\end{equation}
\end{lemma}
\begin{proof}
By the equations (8.81), (8.82) of \cite{AKKR}, we get
\begin{equation} \label{6}
\beta \eta f \left(\frac{\beta}{4 \mathfrak{m} \eta} \right) \leq
\hat{u}_0 .
\end{equation}
As in \cite{pk}, we use the Bogolyubov inequality
\[
\frac{\beta}{2} \varrho_{\beta,\Lambda_{0,l}}  \left\{ A A^* + A^*
A\right\} \cdot \varrho_{\beta,\Lambda_{0,l}}\left\{ \left[ C^* ,
\left[H, C\right]\right]\right\} \geq \left\vert
\varrho_{\beta,\Lambda_{0,l}}\left\{ \left[ C^*, A \right]
\right\} \right\vert^2,
\]
in which we set $A$ to be the identity operator, $C=
\mathfrak{p}_l$, $H = H_{0, l}$, and obtain
\begin{equation} \label{7}
\eta \geq \frac{|a|}{12b}.
\end{equation}
It is not difficult to show that the left-hand side of (\ref{6}) is an increasing function of $\eta$; hence, by
(\ref{7}) one gets the lower bound in (\ref{4}). The upper bound $\hat{u}_0  \leq
\beta \eta$,
 follows from the estimate
(\ref{au34}) (positivity), (\ref{6.120}) and the definition (\ref{1}). One can show (see subsection 4.2 of \cite{AKR}
and subsection 3.2 of \cite{AKPR}) that the measure $\mu_0$ is quasi-invariant with respect to the shifts $\omega
\mapsto \omega + t e_q$, $t\in \mathbb{R}$, $q\in \mathcal{Q}$, where $e_q$ is given by (\ref{base}). Its logarithmic
derivative $\mathfrak{b}_q$ in the direction $e_q$ is
\begin{equation} \label{8}
\mathfrak{b}_q (\omega) = - (\mathfrak{m} q^2 + a) \int_0^\beta
e_q (\tau) \omega(\tau) {\rm d}\tau - 4 b \int_0^\beta e_q (\tau) [\omega (\tau)]^3 {\rm d}\tau.
\end{equation}
This derivative is used in the integration-by-parts formula
\begin{equation} \label{9}
\int_{\Omega} \partial_q f (\omega)  \mu_0 ({\rm d}\omega) = -
\int_{\Omega} f (\omega) \mathfrak{b}_q (\omega)  \mu_0 ({\rm
d}\omega) ,
\end{equation}
where
\[
\partial_q f (\omega) \ \stackrel{\rm def}{=} \
 \left[(\partial/\partial t) f (\omega + t e_q) \right]_{t = 0},
\]
and $f:\Omega \rightarrow \mathbb{R}$ can be taken
\begin{equation} \label{10}
f (\omega )= \int_0^\beta e_q (\tau) \omega(\tau) {\rm d}\tau.
\end{equation}
We apply (\ref{9}) with $q=0$ to the function (\ref{10}), also with $q=0$, and obtain
\begin{equation} \label{11}
1 = - |a| \hat{u}_0  + \frac{ 4b}{\beta}
\int_{\mathcal{I}_\beta^2}\Gamma^{(0)}_4 (\tau, \tau, \tau, \tau')
{\rm d}\tau {\rm d}\tau'.
\end{equation}
By the GKS-inequality (see Theorem 6.2 in \cite{AKKR}),
\[
\Gamma^{(0)}_4 (\tau, \tau, \tau, \tau') \geq \Gamma^{(0)}_2
(\tau, \tau)  \Gamma^{(0)}_2 (\tau, \tau'),
\]
by which and by the estimate $\hat{u}_0  \leq \beta \eta$, we have in (\ref{11})
\[
1 \geq  - |a| \hat{u}_0  + 4 b  \eta \hat{u}_0  \geq  - |a|
\hat{u}_0  + 4 b \beta^{-1}  \hat{u}_0^2,
\]
that is equivalent to the second upper bound in (\ref{4}).

By means of the lattice approximation technique and the estimate (3.15) of \cite{bfs}, one gets
\[
- U_4^{(0)} (\tau_1 , \tau_2 , \tau_3 , \tau_4) \leq 4! b
\int_0^\beta U_2^{(0)} (\tau_1 , \tau) U_2^{(0)} (\tau_2 , \tau)
U_2^{(0)} (\tau_3 , \tau) U_2^{(0)} (\tau_4 ,
\tau){\rm d}\tau,
\]
which yields
\begin{eqnarray*}
X_0 &\leq & 4! b \hat{u}_0^2 \int_{0}^\beta \left[U_2^{(0)} (\tau,
\tau') \right]^2 {\rm d}\tau' \leq 4! b \hat{u}_0^3 \beta \eta
\nonumber \\& \leq & 4! b \hat{u}_0^4 \left[f\left(\frac{\beta}{4
\mathfrak{m} \eta} \right) \right]^{-1},
\end{eqnarray*}
where we have used the upper bound for $\beta \eta$ obtained from (\ref{6}). For $f$ given by (\ref{3}), one can show
that $1/f(t)$ is an increasing function of $t$. Then the estimate (\ref{5}) is obtained from the above one by means of
(\ref{7}).  $\square$
\end{proof}
Let us return to the functions (\ref{h28}), (\ref{au30}). Recall that we suppose $\delta \in (0, 1/2)$. Given $\epsilon
\in \left(0, (1 - 2\delta)/4 \right)$,
 we define ${v} (\epsilon)$ by the
condition $\sigma ({v}(\epsilon)) = \varkappa^\epsilon$. An easy calculation yields
\begin{equation} \label{h27}
{v}(\epsilon) = \frac{\varkappa^\delta - \varkappa^{-\epsilon} }{\varkappa^\delta -1} =
 1 + \frac{1 - \varkappa^{-\epsilon}}{\varkappa^\delta -1}.
\end{equation}
Then
\begin{equation} \label{h30}
\phi (v) \leq \varkappa^{2\delta + 4 \epsilon -1} < 1, \quad
 {\rm for \ } v\in [1, {v}(\epsilon)].
\end{equation}
Furthermore, we set
\begin{equation} \label{au33}
{w}(\epsilon) = 2\varkappa^{1 - \delta - 2\epsilon}\cdot
\frac{(\varkappa^{\delta} - \varkappa^{-\epsilon}) ( 1 -
\varkappa^{- \epsilon})}{(\varkappa^{\delta} - 1)^2},
\end{equation}
\begin{equation} \label{au40}
w_{\rm max} = \sup_{\epsilon \in (0, (1- 2\delta)/4)} w (\epsilon).
\end{equation}
The function $\epsilon \mapsto w (\epsilon)$ is continuous, then for every $w < {w}_{\rm max}$, one finds $\varepsilon
\in (0, (1- 2\delta)/4)$ such that $w < w(\varepsilon )$. Set $\bar{v} = v (\varepsilon )$ and $\bar{w} = w
(\varepsilon)$. Therefore, for this $w$, one has
\begin{equation} \label{au32}
- \psi (v)  w + v \sigma (v) > v, \quad {\rm for \ } v \in [1,
\bar{v}] .
\end{equation}
\begin{lemma} \label{osz1lm}
The parameters $\mathfrak{m}>0$, $a\in \mathbb{R}$ and $b>0$ may be chosen in such a way that there will exist
$\varepsilon \in (0, (1-2\delta)/4)$ and the numbers $\beta_0^{\pm}$, $0< \beta_0^{-} <
\beta_0^{+}< +\infty$ with the following properties: (a)
$\hat{u}_0  = 1$ for $\beta = \beta_0^{-}$ and $\hat{u}_0  < 1$ for $\beta < \beta_0^{-}$; (b) $\hat{u}_0  = \bar{v} =
v(\varepsilon)$ for $\beta = \beta_0^{+}$ and $\hat{u}_0  <
\bar{v}$ for $\beta < \beta_0^{+}$; (c) $X_0 <
\bar{w}=w(\varepsilon)$ for all $\beta \in [\beta_0^{-},
\beta_0^{+}]$.
\end{lemma}
\begin{proof}
 Let us fix $\gamma = |a|/b$. Then by (\ref{4}) and (\ref{3}), one has
\begin{equation} \label{105}
\frac{\mathfrak{m} \gamma^2}{36} \left[1 - \exp\left(-
\frac{3\beta}{\mathfrak{m}\gamma}\right) \right]\leq \hat{u}_0
\leq \frac{\beta \gamma}{8}\left[1 + \sqrt{1+ \frac{16}{\beta
\gamma}} \ \right],
\end{equation}
which immediately yields $\hat{u}_0  \rightarrow 0$ as $\beta
\rightarrow 0$. On the other hand, by taking $\mathfrak{m}
\gamma^2
> 36 \bar{v}$, one gets $\hat{u}_0  > \bar{v}$ for sufficiently
large $\beta$. Since by Lemma \ref{ormlm}, $\hat{u}_0$ depends on $\beta$ continuously, this means that
$\beta_0^{\pm}$, such that $\beta_0^{-} < \beta_0^{+}$, do exist. For fixed $\gamma$ and $\mathfrak{m}$, the multiplier
$[f( 3 \beta /
\mathfrak{m}\gamma)]^{-1}$ in (\ref{5}) is bounded as $\beta \in
(0, \beta_0^{+}]$. Recall, that $\hat{u}_0  \leq \bar{v}$ for such $\beta$. Then, keeping $\gamma$ fixed, we pick up
$b$ such that the right-hand side of (\ref{5}) will be less than $\bar{w}$. $\square$ \vskip.1cm \noindent
\end{proof}
\begin{lemma}
\label{6.3lm} Let $\mathfrak{I}_n $, $n \in \mathbb{N}_0$, be the
triple of statements $(\mathfrak{i}_n^1 , \mathfrak{i}_n^2 ,
\mathfrak{i}_n^3 )$, where
\begin{eqnarray*}
\mathfrak{i}_n^1 & = & \{ \exists \beta_n^{+} \in [\beta_0^- ,
\beta_0^+]  : \ \hat{u}_n  = \bar{v}, \ \beta = \beta_n^{+} ; \
\hat{u}_n
< \bar{v} ,  \ \forall \beta < \beta_n^{+} \};\\
\mathfrak{i}_n^2 & = & \{ \exists \beta_n^{-} \in [\beta_0^- ,
\beta_0^+]  : \ \hat{u}_n  = 1 , \ \beta = \beta_n^{-}; \
\hat{u}_n  < 1 , \ \forall \beta < \beta_n^{-} \}; \nonumber\\
\mathfrak{i}_n^3 & = & \{ \forall \beta \in (0, \beta_n^{+} ) : \
X_n < \bar{w} \}. \nonumber \end{eqnarray*} Then {\rm (i)} $\mathfrak{I}_0 $ is true; {\ } {\rm (ii)}
$\mathfrak{I}_{n-1}$ implies $\mathfrak{I}_n$.
\end{lemma}
\begin{proof} $\mathfrak{I}_0$ is true by
Lemma \ref{osz1lm}. For $\beta = \beta_{n}^{+}$, $\sigma (\hat{u}_n  )  = \varkappa^\varepsilon $ and $\sigma
(\hat{u}_n ) < \varkappa^\varepsilon $ for $ \beta < \beta_n^{+}$ (see (\ref{h27}), (\ref{h30})). Set $ \beta =
\beta_{n-1}^{+}$, then (\ref{au32}), (\ref{au33}), (\ref{6.10}), and $\mathfrak{i}_{n-1}^3$ yield
\begin{eqnarray} \label{6.15}
\hat{u}_n  & \geq & \varkappa^\varepsilon \bar{v}- \frac{1}{2}
(1-\varkappa^{-\delta}) \varkappa^{3 \varepsilon}
 \varkappa^{2\delta -1} X_{n-1}\nonumber\\
& > & \varkappa^\varepsilon  \bar{v} \left[ 1 - \varkappa^{2(\varepsilon - 1)+\delta} (\varkappa^\delta
-1)\frac{\bar{w}}{\bar{v}}\right] =
\bar{v} .
\end{eqnarray} For $\beta = \beta_{n-1}^{-}$, the estimate (\ref{6.8})
gives
\begin{equation} \label{6.16}
 \hat{u}_n <
1.
\end{equation}
Taking into account Lemma \ref{ormlm} (continuity) and the estimates (\ref{6.15}), (\ref{6.16}), one concludes that
there exists at least one value $\tilde{\beta}_n^{+} \in (\beta_{n-1}^{-}, \beta_{n-1}^{+})$ such that $\hat{u}_n  =
\bar{v} $. Then we put $\beta_n^+ = \min\tilde{\beta}_n^+$. The
mentioned continuity of $\hat{u}_n$ yields also $\hat{u}_n<\bar{ v}$ for $\beta < \beta_n^+$. Thus $\mathfrak{i}_n^1 $
is true. The existence of $ \beta_n^{-} \in [\beta_{n-1}^{-}, \beta_{n-1}^{+})$ can be proven in the same way. For
$\beta < \beta_n^{+} <
\beta_{n-1}^{+}$, we have $\sigma (\hat{u}_{n -1}) <
\varkappa^\varepsilon $, which yields
\begin{equation} \label{6.17}
X_n  < \varkappa^{2\delta -1} \varkappa^{4 \varepsilon } X_{n-1}  \leq  X_{n-1} < \bar{w },
\end{equation}
 hence, $\mathfrak{i}_n^3 $ is true as well.
 The proof is concluded by
remarking that
\begin{equation} \label{6.18}
[\beta_n^{-}, \beta_n^{+}] \subset [\beta_{n-1}^{-},
\beta_{n-1}^{+}) \subset [\beta_0^{-}, \beta_0^{+}]
.\end{equation} $\square$
\end{proof}
\begin{lemma}
\label{Newlm} There exists $\beta_* \in [\beta_0^{-},
\beta_0^{+}]$ such that, for $\beta = \beta_*$, the following
estimates hold for all $n\in \mathbb{N}_0$:
\begin{equation}
\label{new} 1 < \hat{u}_n  < \bar{v}.
\end{equation}
For $\beta < \beta_* $, the above upper estimate, as well as the estimate(\ref{neww}), hold.
\end{lemma}
\begin{proof}
 Consider the set $\Delta _{n}\ \stackrel{\rm def}{=}
\{\beta \in (0,{\beta }_{n}^{+}) \ | \ 1 < \hat{u}_{n} <
\bar{v}\}$. Just above we have shown that it is nonempty and
$\Delta _{n}\subseteq ({\beta }_{n}^{-},{\beta }_{n}^{+} )$. Let us prove that $\Delta _{n}\subseteq \Delta _{n-1}$.
Suppose there exists some $\beta \in \Delta _{n}$, which does not belong to $\Delta _{n-1}$. For this $\beta $, either
$\hat{u}_{n-1}  \leq 1$ or $\hat{u }_{n-1} \geq \bar{v}$. Hence, either $\hat{u}_{n} <1$ or $\hat{ u}_{n} > \bar{v}$
(it can be proven as above), which is in conflict with the assumption $\beta \in \Delta _{n}$. Now let $D_n$ be the
closure of $\Delta_n$, then one has
\begin{equation} \label{newv}
D_n = \{ \beta \in [\beta_n^- , \beta_n^+ ]\ | \ 1\leq \hat{u}_n  \leq v(\delta) \},
\end{equation}
which is a nonempty closed set. Furthermore, $D_n \subseteq D_{n-1} \subseteq \dots \subset [\beta_0^- , \beta_0^+ ]$.
Set $D_{\ast }=\bigcap_{n}D_{n}$, then $D_{\ast }\subset [\beta_0^{-},
\beta_0^{+}]$ is also nonempty and closed. Now let us show that,
for every $\beta \in D_{\ast}$, the sharp  bounds in (\ref{new}) hold for all $n\in \mathbb{N}$. Suppose $\hat{u}_n  =
\bar{v}$ for some $n\in \mathbb{N}$. Then (\ref{6.15}) yields $\hat{u}_m  >
\bar{v}$ for all $m>n$, which means that this $\beta$ does not
belong to all $D_m $, and hence to $D_{\ast}$. Similarly one proves the lower bound by means of (\ref{6.8}). On the
other hand, by means of the above arguments, one can conclude that $\beta\in D_*$ if the inequalities (\ref{new}) hold
for all $n\in
\mathbb{N}_0$ at this $\beta$. Set $\beta_{\ast }=\min D_{\ast }$.
Then (\ref{new}) hold for $\beta =\beta _{\ast }$. Let us prove (\ref{neww}). Take $\beta <\beta _{*}.$ If $\hat{u}_{n}
>1$ for all $ n\in \mathbb{N}$, then either (\ref{new}) holds or
there exists such $n_{0}\in \mathbb{N}$ that $\hat{u}_{n_{0}} \geq
\bar{v}$. Therefore, either $\beta \in D_{*}$ or $\beta >\inf
{\beta }_{n}^{+}.$ Both these cases contradict the assumption $\beta < \beta _{*}$. Thus, there exists $ n_{0}\in
\mathbb{N}$ such that $\hat{u }_{n_0 -1 } \leq 1$ and hence $\hat{u}_n < 1$ for all $n\geq n_{0}.$ In what follows, the
definition (\ref {h28}) and the estimate (\ref{6.8}) imply that the sequences $\{\hat{u} _{n}
 \}_{ n\geq n_{0}}$ and $\{\sigma(\hat{u}_{n} ) \}_{ n\geq
n_{0}} $ are strictly decreasing. Then for all $n>n_{0}$, one has (see (\ref{6.8}))
\begin{eqnarray}
\hat{u}_{n}  & < & \sigma (\hat{u}_{n-1} ) \hat{u}_{n-1}  < \dots
\nonumber \\ & < & \sigma(\hat{u}_{n-1} ) \sigma (\hat{u}_{n-2} )
\dots \sigma (\hat{u}_{n_0} )\hat{u}_{n_0}
 < \left[\sigma (\hat{u}_{n_0})\right]^{n-n_{0}}. \nonumber
\end{eqnarray}
Since $\sigma (\hat{u}_{n_{0}})<1,$ one gets $\sum_{n=0}^{\infty }
\hat{u}_{n} <\infty.$ Thus,
\[
\prod_{n=1}^{\infty } \left[1-(1-\varkappa^{-\delta})\hat{u}_{n
-1} \right]^{-1}
 \stackrel{\rm def}{=}K_{0} < \infty .
\]
Finally, we apply (\ref{6.8}) once again and obtain
\begin{eqnarray*}
\hat{u}_{n}  & < & \varkappa^{-n\delta}
\left[1-(1-\varkappa^{-\delta})\hat{u}_{n-1} \right]^{-1} \dots
\left[1-(1-\varkappa^{-\delta})\hat{u}_{0} \right]^{-1} \hat{u}_0  \\
& < & \varkappa^{-n \delta}K_0 \bar{v} \stackrel{\rm def}{=}  K (\beta )\varkappa^{-n \delta} .
\end{eqnarray*}
$\square$
\end{proof}
\vskip.1cm \noindent
 {\it Proof of Lemma \ref{5.5lm}}. The existence of $\beta_*$ has
 been proven in Lemma \ref{Newlm}.
Consider the case $\beta = \beta_* $ where the estimates (\ref{new}) hold. First we show that $X_n \rightarrow 0$.
Making use of (\ref{6.11}) we obtain
\[
 0 < X_n \leq
\varkappa^{2\delta - 1}\left[\sigma (\hat{u}_{n-1})\right]^4
X_{n-1} < X_{n-1} < X_{n-2} < \cdots < \bar{w}.
\]
Therefore, the sequence $\{X_n \}$ is strictly decreasing and bounded, hence, it converges and its limit, say $X_* $,
obeys the condition $X_* < X_0 <
 \bar{w}$. Assume that
$X_* >0$. Then (\ref{6.11}) yields $\sigma (\hat{u}_n )
\rightarrow \varkappa^\varepsilon $ hence $\hat{u}_n  \rightarrow
\hat{u}_\infty  \geq \bar{v}$. Passing to the limit $n\rightarrow
\infty$ in (\ref{6.10}) one obtains $X_* \geq \bar{w}$ which
contradicts the above condition. Thus $X_* = 0$. To show $\hat{u}_n  \rightarrow 1$ we set
\begin{equation} \label{newx}
\Xi_n = - \frac{1}{2}
(1-\varkappa^{-\delta})\left[\sigma(\hat{u}_{n-1} )\right]^3
\varkappa^{2\delta -1} X_{n-1}.
\end{equation}
Combining (\ref{6.8}) and (\ref{6.10}) we obtain
\begin{equation} \label{newy}
0\geq \hat{u}_n
 - \sigma (\hat{u}_{n-1}  ) \hat{u}_{n-1}
 \geq \Xi_n \rightarrow 0 .
\end{equation}
For $\beta = \beta_*$, we have $\{\hat{u}_n  \}\subset [1,
\bar{v})$ in view of Lemma \ref{Newlm}. By (\ref{newy})  all its
accumulation points in $\in [1, \bar{v}]$  ought to solve the equation $$ u- \sigma (u) u =0. $$ There is only one such
point: $u_* =1$, which hence is the limit of the whole sequence $\{\hat{u}_n  \}$. For $\beta <\beta_* $, the estimate
(\ref{neww}) has been already proven in Lemma \ref{Newlm}. This yields $\sigma (\hat{u}_n  ) \rightarrow
\varkappa^{-\delta}$, which implies $X_n \rightarrow 0$ if (\ref{6.11}) is taken into account.

  $\square$

\end{document}